\documentclass[cernpreprint, atlasdraft=false, texlive=2023, UKenglish, texmf, orcidlogo]{atlasdoc}
 
\usepackage{atlaspackage}
\usepackage{atlasbiblatex}
 
\usepackage{atlasphysics}
\usepackage{atlashepprocess}
\usepackage{multirow}

\graphicspath{{logos/}{figures/}}
 
\usepackage{ANA-EXOT-2022-14-PAPER-defs}

\AtlasTitle{Search for top-philic heavy resonances in $pp$ collisions at $\sqrt{s}=13$~\tev\ with the ATLAS detector}
 
\AtlasAbstract{
A search for the associated production of a heavy resonance with a top-quark or a top-antitop-quark pair, and decaying into a $t\bar{t}$ pair is presented.
The search uses the data recorded by the ATLAS detector in $pp$ collisions
at $\sqrt{s}= 13$~\tev\ at the Large Hadron Collider during the years 2015--2018,
corresponding to an integrated luminosity of 139~\ifb.
Events containing exactly one electron or muon are selected.
The two hadronically decaying top quarks from the resonance decay are reconstructed using jets clustered with a large radius parameter of $R=1$.
The invariant mass spectrum of the two top quark candidates is used to search for a resonance signal in the range of 1.0~\tev\ to 3.2~\tev.
The presence of a signal is examined using an approach with minimal model dependence followed by a model-dependent interpretation.
No significant excess is observed over the background expectation.
Upper limits on the production cross section times branching ratio at 95\% confidence level are provided
for a heavy $Z^\prime$ boson based on a simplified model,
for $Z^\prime$ mass between 1.0~\tev\ and 3.0~\tev.
The observed (expected) limits range from 21~(14)~fb to 119~(86)~fb depending on the choice of model parameters.
 
}
 
\author{The ATLAS Collaboration}
 
\AtlasRefCode{EXOT-2022-14}
 
\AtlasJournalRef{\href{https://link.springer.com/article/10.1140/epjc/s10052-023-12318-9}{Eur. Phys. J. C 84 (2024) 157}}
\AtlasDOI{10.1140/epjc/s10052-023-12318-9}
 
\PreprintIdNumber{CERN-EP-2023-048}

\AtlasHeading{ATLAS Paper Draft}

\hypersetup{pdftitle={ATLAS document},pdfauthor={The ATLAS Collaboration}}
 
\begin{document}
 
\maketitle
 
\tableofcontents

% The next lines are included from the .//intro/intro.tex input file
\section{Introduction}
\label{sec:introduction}
 
\begin{figure}[bh!]
\centering
\subfloat[]{
\label{fig:introd:BSM4tops}
\includegraphics[width=0.30\linewidth]{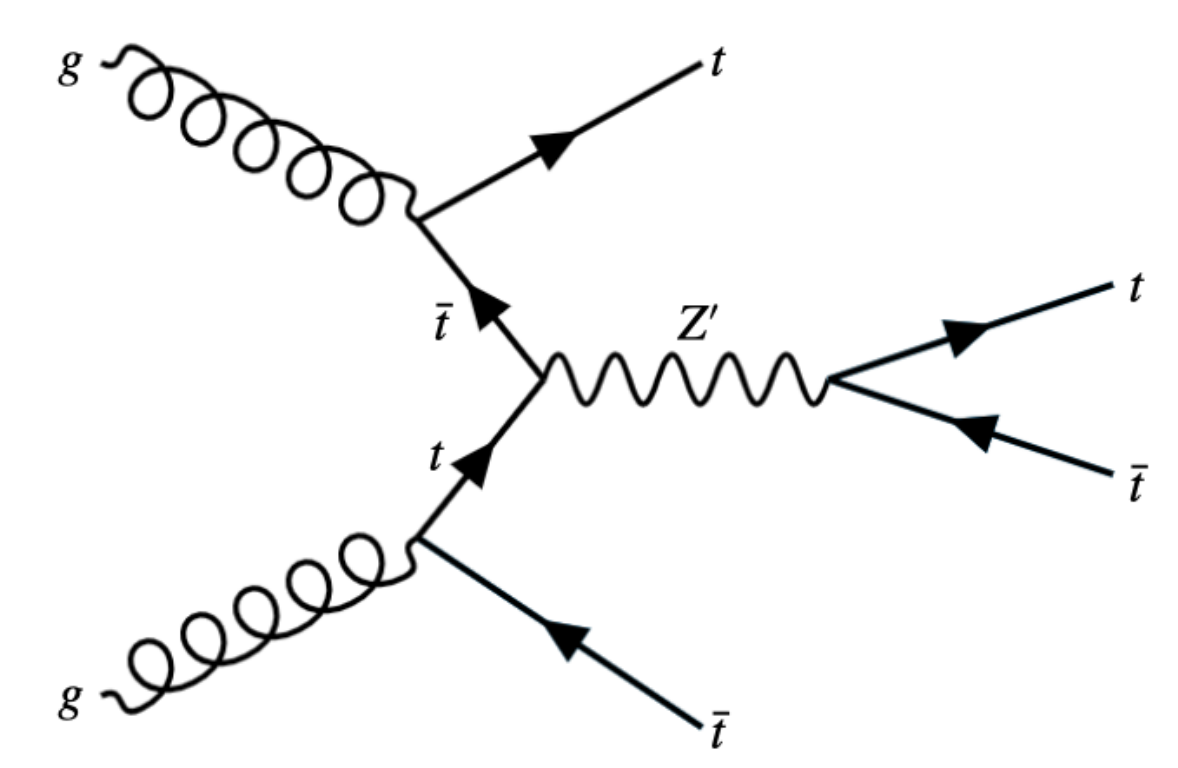}
}
\subfloat[]{
\label{fig:introd:tJZPrime}
\includegraphics[width=0.30\linewidth]{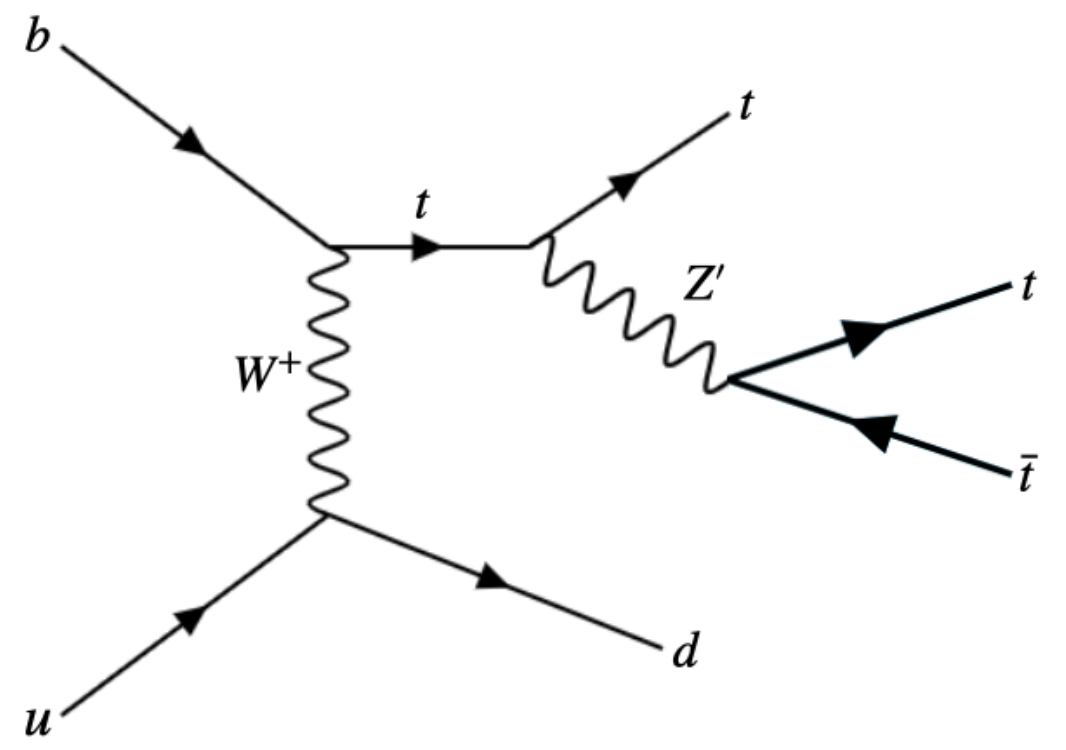}
}
\subfloat[]{
\label{fig:introd:tWZPrime}
\includegraphics[width=0.30\linewidth]{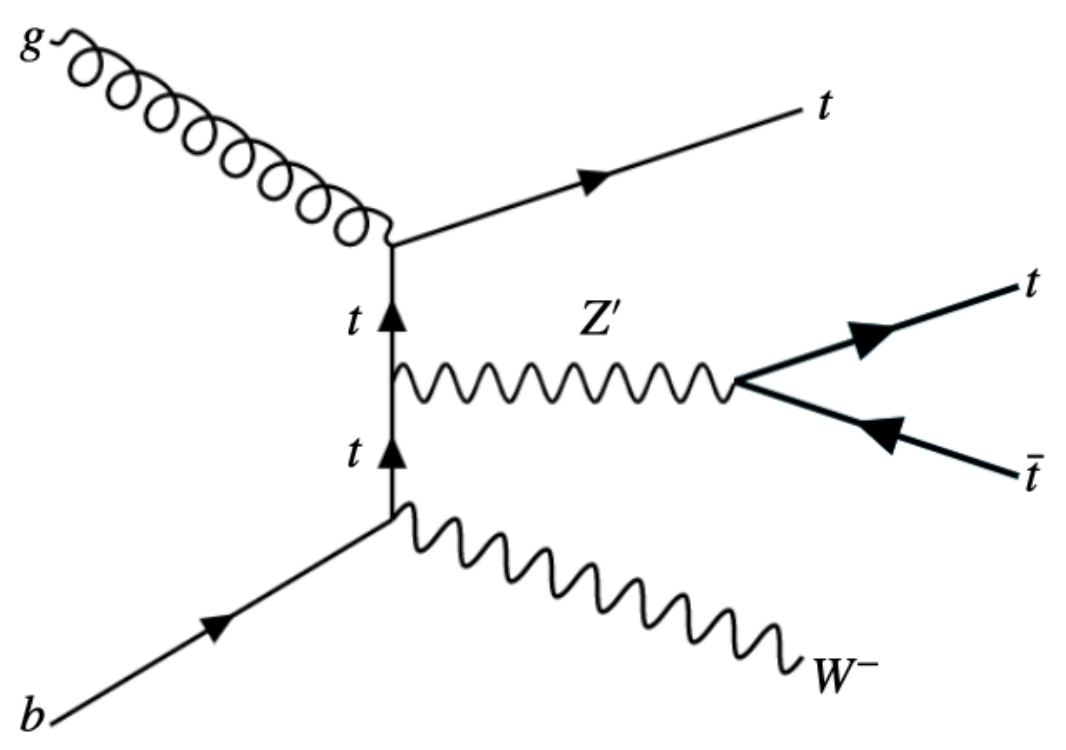}
}
 
\caption{Examples of tree-level Feynman diagrams for \Zprime production in association with (a) $\ttbar$, (b) $tj$ (where $j$ refers to any light quark), and (c) $tW$. The \Zprime generation modes are derived from top quark final states produced via (a) strong, (b) electroweak, and (c) mixed QCD and electroweak interactions.}
\label{fig:introd:signal}
\end{figure}

The discovery of a new particle consistent with the Standard Model (SM)
Higgs boson by the ATLAS \cite{HIGG-2012-27} and CMS \cite{CMS-HIG-12-028}
collaborations was a major milestone in high-energy physics. However, the
underlying nature of electroweak symmetry breaking remains unknown.
Naturalness arguments suggest that the large quantum corrections to
the Higgs boson mass are cancelled by some new mechanism to prevent an excessive level of fine-tuning.
Such a mechanism has been proposed in several theories of Beyond the SM (BSM) physics.
The large Yukawa coupling of the top quark to the Higgs boson motivates various top-quark-based resonance searches.
In many BSM theories, such as composite Higgs scenarios~\cite{Ferretti:2013kya,Vecchi:2015fma,Agashe:2003zs,Agashe:2004rs},
new `top-philic' vector resonances are predicted.
These resonances couple more strongly to the top quark than to the light quarks,
such that all the other couplings except for the one with top quarks can be neglected.
Typical \ttbar\ resonance searches target new resonances produced through \qqbar\ annihilation,
assuming sizeable couplings to light quarks~\cite{EXOT-2016-24,EXOT-2015-04,CMS-B2G-16-015,CMS-B2G-17-017}.
Top-philic resonances, on the other hand, require different production modes, such as
the production of the heavy resonance in association with a top-quark or a \ttbar\ pair,
resulting in three or four top quarks in the final state.
\Fig{\ref{fig:introd:signal}} shows tree-level diagrams for the production of a top-philic resonance $Z'$.
 
This analysis searches for a new top-philic heavy resonance above a mass of 1.0~\tev.
The ATLAS and CMS collaborations
have previously published measurements of SM four-top-quark (\tttt) production~\cite{TOPQ-2018-05,TOPQ-2020-10,CMS-TOP-18-003,Sirunyan:2019nxl,CMS-TOP-21-005}.
with the first observation of this process reported by both experiments recently~\cite{TOPQ-2021-08,CMS-TOP-22-013}.
Three-top-quark production, with a much smaller cross section of $\mathcal{O}$(1~fb) predicted by the SM~\cite{Barger:2010uw,Chen:2014,Boos:2021yat}, has not been measured.
In resonant BSM models, the two top quarks from the resonance decay, referred to as the `resonance top quarks', are expected to be highly boosted.
The other top quarks are referred to as the `spectator top quarks' and are expected to have lower momenta.
While previous BSM searches explored the \tttt final state for several target models~\cite{EXOT-2016-13,EXOT-2016-16},
this analysis uniquely attempts to reconstruct the resonance explicitly,
enabling a search of a new physics contribution with minimal model dependence.
 
In addition,
a simplified model~\cite{Greiner:2014qna} considering a colour-singlet vector particle \Zprime
is used to generate simulated samples for further model-dependent interpretations, however without relying on how the resonance couples to other particles in a specific model. The interaction Lagrangian reads:
\begin{align}
\nonumber
\mathcal{L} &= \overline{t} \gamma_{\mu} \left(c_{L} P_{L} + c_{R} P_{R}\right) t Z^{\prime\mu} \\ \nonumber
&= c_t \overline{t} \gamma_{\mu} \left(\cos{\theta} P_{L} + \sin{\theta} P_{R}\right) t Z^{\prime\mu},
\end{align}
where $\gamma_{\mu}$ are the Dirac matrices and \(P_{L/R} = (1 \mp \gamma_{5}) / 2\) are the chirality projection operators with $\gamma_5=i\gamma^0\gamma^1\gamma^2\gamma^3$. The coupling of the
vector singlet to the top quarks is defined as \(c_t = \sqrt{c_L^2 + c_R^2}\), with components that couple only to left-handed and right-handed top-quarks $c_L$ and $c_R$, respectively. The tangent of the chirality angle is defined as \(\tan{\theta} = c_R / c_L\).
For the mass range which is explored in this search with a $Z'$ mass ($\mZp$) much larger compared to the top-quark mass ($m_t$), the $Z'$ decay width can be approximated by $\Gamma/\mZp \approx c_t^2/(8\pi)$. To minimise model dependence, only the tree-level production of a heavy top-philic resonance is considered. At the LHC a large contribution comes from \ttZp\ (\Fig{\ref{fig:introd:BSM4tops}}) which is independent of $\theta$. Further contributions arise from $tjZ'$ (\Fig{\ref{fig:introd:tJZPrime}}) and $tWZ'$ (\Fig{\ref{fig:introd:tWZPrime}}) production modes, which are negligible for $\theta =\pi/2$, and are largest for $\theta = 0$ with similar magnitude as \ttZp.

This search uses the data recorded by the ATLAS detector in \pp collisions at
$\sqrt{s}=13$~\tev\ between 2015 and 2018, corresponding to 139~\ifb.
In the SM, the top quark is expected to decay into a $W$ boson and a $b$-quark with a branching ratio of approximately 100\% with subsequent hadronic or leptonic decay of the $W$ boson.
This analysis targets final states for which both resonance top quarks decay hadronically and one of the spectator top quarks decays leptonically. The single-lepton channel is preferred over the fully hadronic channel due to the significantly lower background arising from multijet processes, and the ability to trigger on events with at least one lepton.
The top quarks from the resonance decay are expected to be highly boosted, and therefore their hadronic decays are reconstructed using jets with a large radius parameter and identified by requiring the jets to have a large mass and momentum.
This final state with one lepton\footnote{In the rest of this article, `lepton' refers exclusively to an electron or a muon.
Tau leptons are not explicitly considered although their decay products can be accepted by the electron, muon and jet selection criteria.} suffers from a large background that is mostly composed
of \ttbar\ production in association with additional jets (\ttjets), especially when the associated jets contain $b$-hadrons.
A data-driven technique assisted by simulations is used to overcome the challenge of modelling the \ttjets background.
The presence of a signal is tested in signal-enriched regions
using an approach with minimal model dependence followed by a model-dependent interpretation.

% End of text imported from the .//intro/intro.tex input file

% The next lines are included from the .//atlas/atlas.tex input file
\section{The ATLAS detector}
\label{sec:atlas}
 
The ATLAS detector~\cite{PERF-2007-01} at the LHC is a multipurpose particle detector
with a forward--backward symmetric cylindrical geometry with a nearly \(4\pi\) coverage in
solid angle.\footnote{ATLAS uses a right-handed coordinate system with its origin at the nominal interaction point (IP) in the centre of the detector and the \(z\)-axis along the beam pipe. The \(x\)-axis points from the IP to the centre of the LHC ring, and the \(y\)-axis points upwards. Cylindrical coordinates \((r,\phi)\) are used in the transverse plane, \(\phi\) being the azimuthal angle around the \(z\)-axis. The pseudorapidity is defined in terms of the polar angle \(\theta\) as \(\eta = -\ln \tan(\theta/2)\). Angular distance is measured in units of \(\Delta R \equiv \sqrt{(\Delta\eta)^{2} + (\Delta\phi)^{2}}\).}
It consists of an inner tracking detector (ID) surrounded by a superconducting solenoid,
electromagnetic (EM) and hadron calorimeters, and a muon spectrometer (MS).
 
The ID covers the pseudorapidity range \(|\eta| < 2.5\).
The high-granularity silicon pixel detector covers the vertex region and typically provides four measurements per track, the first hit normally being in the insertable B-layer (IBL) installed before Run 2~\cite{ATLAS-TDR-19, PIX-2018-001}.
It is surrounded by a silicon microstrip detector and a straw tube transition-radiation tracking detector.
The calorimeter system covers the pseudorapidity range \(|\eta| < 4.9\).
It consists of lead/liquid-argon (LAr) sampling calorimeters which provide EM energy measurements
with high granularity.
A steel/scintillator-tile hadron calorimeter covers the central pseudorapidity range (\(|\eta| < 1.7\)).
The endcap and forward regions are instrumented with LAr calorimeters
for EM and hadronic energy measurements up to \(|\eta| = 4.9\).
The MS surrounds the calorimeters and is based on
three large air-core toroidal superconducting magnets with eight coils each.
The field integral of the toroids ranges between \num{2.0} and \SI{6.0}{\tesla\metre} across most of the detector.
The MS includes a system of precision tracking chambers and fast detectors for triggering.
The ATLAS trigger and data acquisition system~\cite{TRIG-2016-01} consists of a two-level trigger system in order to select events.
The first-level trigger is implemented in hardware and uses a subset of the detector information
to accept events at a maximum rate of nearly \SI{100}{\kHz}.
The second-level is a software-based trigger that
reduces the accepted event rate to \SI{1}{\kHz}, on average,
depending on the data-taking conditions~\cite{TRIG-2016-01}.
An extensive software suite~\cite{ATL-SOFT-PUB-2021-001} is used for real and simulated data reconstruction
and analysis, for operation and in the trigger and data acquisition systems of the experiment.

% End of text imported from the .//atlas/atlas.tex input file

% The next lines are included from the .//selection/selection.tex input file
\section{Object reconstruction and event selection}
\label{sec:selection}
 
This analysis uses a set of data events collected by the ATLAS detector between 2015 and 2018 at
$\sqrt{s}=13$~\tev .
Only events for which all detector subsystems were operational are considered.
The data set corresponds to an integrated luminosity of 139~\ifb~\cite{ATLAS-CONF-2019-021,LUCID2}.
 
This analysis is based on events where the detector read-out is triggered by the presence of at
least one electron or one muon, referred to as single-lepton triggers.
Events are selected if the leptons satisfy either low transverse momentum \pt thresholds,
with identification and isolation requirements, or a looser identification criterion with
no isolation requirement and higher \pt thresholds. The lowest \pt requirement used in the single-lepton triggers varies
from 20 to 26 \GeV\ depending on the data-taking period and the lepton flavour~\cite{TRIG-2018-05,TRIG-2018-01}.
 
Events are required to have at least one primary vertex reconstructed from at least two ID tracks with
\(\pt > 500~\MeV\). For events with several primary vertices, the one with the largest sum of the squared
transverse momenta of the associated tracks is taken~\cite{ATL-PHYS-PUB-2015-026}.
 
Electron candidates are reconstructed from an isolated electromagnetic calorimeter
energy deposit which is matched to a track in the ID~\cite{EGAM-2018-01}.
The pseudorapidity of the calorimeter energy cluster, $\eta_{\text{cluster}}$, must satisfy $|\eta_{\text{cluster}}|<2.47$,
excluding the transition region between the barrel and the end-cap EM calorimeter (${|\eta_{\text{cluster}}|\not\in[1.37,1.52]}$).
Muon candidates are reconstructed by combining tracks in the ID with tracks in
the MS~\cite{PERF-2015-10} and are required to have $|\eta|<2.5$. Both electron and muon candidates
are required to have a \pt above 28~\GeV. Further selections on the longitudinal and transverse impact parameters
are imposed. The transverse impact parameter divided by its estimated uncertainty, ${|\dzero|/\sigma(\dzero)}$,
has to be less than five (three) for electron (muon) candidates. The longitudinal impact parameter $z_0$
is required to be ${|\zzsth|<0.5}$~mm for both electron and muon candidates.
Electron candidates must pass a `Tight' likelihood-based identification working point~\cite{EGAM-2018-01}
employing calorimeter and tracking information that provides separation between electrons and jets.
They are also required to be isolated according to the `Tight' selection, a criteria
based on the properties of the topological clusters in the calorimeter and of
the ID tracks around the reconstructed electron~\cite{EGAM-2018-01}.
Muon candidates must satisfy the `Medium' cut-based identification working point~\cite{Aad:2020gmm}.
They are also required to be isolated using the `TightTrackOnly' selection, a criteria based on the properties of the ID tracks around the reconstructed
muon~\cite{Aad:2020gmm}.
 
Jets are reconstructed from topological clusters of calorimeter cells and tracking information
using the \antikt algorithm~\cite{Cacciari:2008gp,Fastjet} with a radius parameter of \(R=0.4\),
processed using a particle-flow algorithm~\cite{PERF-2015-09}.
They are referred to as `small-$R$ jets'.
Jets are required to have \(\pt > \SI{25}{\giga\electronvolt}\) and \(|\eta| < 2.5\) and
are calibrated as described in Ref.~\cite{JETM-2018-05}.
The Jet Vertex Tagger (JVT) discriminant \cite{ATLAS-CONF-2014-018} is used to identify
jets originating from the hard-scatter interaction through the use of tracking and
vertexing information. Jets with \(\pt < \SI{60}{\giga\electronvolt}\) and
\(|\eta| < 2.4\) are required to satisfy the requirement \(\text{JVT} > 0.50\),
corresponding to a selection efficiency for hard-scatter jets of about \SI{96}{\percent}~\cite{PERF-2014-03}.
The DL1r classification algorithm based on recurrent neural networks~\cite{ATLAS:2022qxm,
ATLAS-CONF-2018-001} is used to identify jets containing $b$-hadrons (\(b\)-jets).
This algorithm identifies \(b\)-jets against backgrounds of light-flavour and charm-quark-initiated
jets using information about the impact parameters of tracks associated with the jet
and the topological properties of the displaced vertices reconstructed within the jet.
This analysis uses a \(b\)-jet efficiency working point of \SI{77}{\percent} as measured
for jets with \(\pT > 20~\GeV\) and $|\eta| < 2.5$ in simulated \ttbar events~\cite{FTAG-2018-01}.
The tagging algorithm gives an expected rejection factor (defined as the inverse of the efficiency) of about 192 against light-flavour jets,
and about 5.6 against jets originating from charm quarks~\cite{ATLAS-CONF-2018-001,FTAG-2019-02}.
 
To resolve the potential ambiguities of a single detector response being assigned to two
objects by the reconstruction algorithm, a sequential overlap removal procedure is applied.
First, any electron found to share a track with another electron with higher \pT is removed. Second, electrons sharing their track with a muon candidate are removed. Third, the closest jet in rapidity $y$ and in $\phi$ using $\Delta R_y = \sqrt{(\Delta y)^2+(\Delta \phi)^2} = 0.2$ of
an electron is removed.\footnote{The rapidity is defined as $y=\frac{1}{2} \ln{\frac{E+p_z}{E-p_z}}$,
where $E$ is the energy and $p_z$ is the longitudinal component of the momentum along the beam pipe.}
Fourth, the electrons within $\Delta R_y = 0.4$ of a remaining jet are removed.
Fifth, the jets with less than three associated tracks that are within $\Delta R_y = 0.2$ of a muon are removed.
Finally, remaining muons are removed if their track is within $\Delta R_y = \min( 0.4, 0.04 + \SI{10}{\GeV}/{\pT}_{\mu} )$ of a remaining jet.
 
After the overlap removal,
the selected and calibrated small-$R$ jets are used as inputs for jet reclustering~\cite{ATLAS-CONF-2017-062}
using the \antikt algorithm with a radius parameter of $R=1$.
These reclustered jets are referred to as `large-$R$ jets',
and are used as proxies of the hadronically decaying top quarks in this analysis.
The calibration corrections and uncertainties for the reclustered large-$R$ jets are inherited from the
small-$R$ jets~\cite{ATLAS-CONF-2017-062}.
A trimming procedure~\cite{Krohn:2009th} is applied to reclustered large-R jets which removes all the
associated small-$R$
jets that have a \pT\ below 5\% of the \pT of the reclustered jet to suppress gluon radiation and mitigate pile-up effects.
The large-$R$ jets are required to have \(\pt > \SI{300}{\giga\electronvolt}\), \(|\eta| < 2\), mass \(m > \SI{100}{\giga\electronvolt}\),
and at least two constituent jets.
 
At preselection level, the events are required to have exactly one lepton with $\pT>\SI{28}{\GeV}$ that
matches the lepton that fired the trigger, and at least two large-$R$ jets to capture the two resonance top quarks that decay hadronically.
Additional jets are expected from the decay of the spectator top quarks for \ttZp\ signal events, as well as from the associated production in the $tjZ'$ and $tWZ'$ production modes.
The number of additional jets is counted using small-$R$ jets with \(\Delta R > 1\) to any of the selected large-$R$ jets in an event.
Events are further required to have at least two additional small-$R$ jets and at least two \(b\)-tagged small-$R$ jets.
The $b$-tagged jets can be either inside or outside of the large-$R$ jets and therefore can also count towards additional jets.

% End of text imported from the .//selection/selection.tex input file

% The next lines are included from the .//samples/samples.tex input file
\section{Event simulation}
\label{sec:samples}
 
Monte Carlo (MC) samples of simulated events were produced to model the
signal and background processes. The samples
are normalised using the best theory predictions available.
The generated event samples are processed through the full ATLAS detector~\cite{SOFT-2010-01}
based on \textsc{Geant4}~\cite{Agostinelli:2002hh}. Only the MC samples describing SM \tttt\ production and systematic variations of the \ttbar simulation are processed  through a faster simulation making use of parameterised
showers in the calorimeters~\cite{ATL-PHYS-PUB-2010-013}.
Additional simulated \pp collisions generated
using \PYTHIA{8.186}~\cite{Sjostrand:2007gs} using the \textsc{A3} set of tuned parameters (tune)~\cite{ATL-PHYS-PUB-2012-003}
and the \texttt{MSTW2008LO} parton distribution function (PDF) set~\cite{Martin2009} were overlaid to model the effects of multiple interactions in the same and nearby bunch crossings (pile-up). The distribution of the number of
additional \pp interactions in the MC samples is re-weighted to match the one observed in data. All
simulated samples were processed through the same reconstruction algorithms and analysis chain as the data.
For all samples of simulated events, except those generated using \SHERPA~\cite{Gleisberg:2008ta}, the \EVTGEN~1.2.0
program~\cite{Lange:2001uf} was used to describe the decays of bottom and charm hadrons.
The \texttt{NNPDF3.0NLO}~\cite{Ball:2014uwa} PDF set was used in all
matrix element (ME) calculations if not stated otherwise.
 
The signal samples are simulated using the \MGNLO~2.8.1~\cite{Alwall:2014hca} generator at leading-order (LO) in the five-flavour scheme with the
\texttt{NNPDF3.1LO}~\cite{Ball:2014uwa} PDF set.
The model implementation is based on the simplified top-philic resonance of Ref.~\cite{PhysRevD.94.035023}
where a colour singlet vector particle is considered to exclusively couple to top and anti-top quarks.
Six mass points are generated, spaced according to the experimental resolution.
The \Zprime masses are: \SI{1.0}{\tera\electronvolt},
\SI{1.25}{\tera\electronvolt}, \SI{1.5}{\tera\electronvolt}, \SI{2.0}{\tera\electronvolt},
\SI{2.5}{\tera\electronvolt}, and \SI{3.0}{\tera\electronvolt}.
Interference effects between \(s\)- and \(t\)-channel processes for \ttZp\ production are simulated. The interference with SM four-top-quark production was found to be negligible.
Only the resonant $Z'$ production mode is included in the simulation
of the \(tW Z'\) and \(tj Z'\) channels.
The signal samples are generated with baseline choices of the chirality parameter \(\theta = \pi / 4\) and the coupling between \Zprime and top quarks of \(c_{t} = 3\). Other choices of these parameters are realised using \MADGRAPH matrix element reweighting~\cite{Mattelaer:2016gcx}.
The width of the resonance is computed automatically using \MGNLO and corresponds to a relative width of
about \SI{4}{\percent} for $c_t = 1$ and increases to \SI{64}{\percent} for $c_t = 4$.
 
The production of \ttbar events is modelled using the HVQ program~\cite{Frixione:2007nw,Frixione:2007nu}
in the \POWHEGBOX~2~\cite{Frixione:2007nw,Nason:2004rx,Frixione:2007vw,Alioli:2010xd} generator at next-to-leading order (NLO) in QCD.
The \(h_{\textnormal{damp}}\)
parameter, which controls the transverse momentum \pt
of the first additional emission beyond the Born configuration, was set to 1.5~\(m_{\textnormal{t}}\)~\cite{ATL-PHYS-PUB-2016-020}.
 
The \ttjets\ MC events are classified according to the flavour of the particle jets.
The particle jets are reconstructed from the simulated stable particles
using the anti-\kt algorithm with a radius parameter $R=0.4$,
and are required to have $\pt>15$~\GeV\ and $\abseta<2.5$.
Events are labeled as \ttbjets\ if at least one particle jet is matched within $\Delta R < 0.3$ to
any $b$-hadron with $\pt>5$~\GeV. In the remaining events, if at least one particle jet is matched within $\Delta R<0.3$ to any $c$-hadron with $\pt>5$~\GeV,
the events are labeled as \ttcjets. Only hadrons not associated with $b$- and $c$-quarks from top-quark and $W$-boson decays are considered.
All other events are labeled as \ttljets.
Events categorised as \ttbjets\ and \ttcjets\ are collectively referred to as $\ttbar+\text{heavy flavour}$ (HF) events.

Samples of single-top quark proudction backgrounds corresponding to the \(Wt\) associated production,
\(s\)-channel and \(t\)-channel production, were modelled using the \POWHEGBOX~2
generator at NLO in QCD.
Overlaps between the \ttbar and \(Wt\) final states are removed using the `diagram removal' scheme~\cite{Frixione:2008yi}.
 
A sample of \(\ttbar+H\) events is modeled using the \POWHEGBOX generator at NLO.
The production of \(\ttbar + V\) (with \(V = W, Z\), including non-resonant \(Z/\gamma^{*}\) contributions
and \(\ttbar+WW\)) events are modeled using the \MGNLO~{2.3.3} generator at NLO.
Samples of \Vjets (\(V=W, Z\)) events are generated with the
\SHERPA~{2.2.1}~\cite{Bothmann2019} generator using NLO-accurate MEs for
up to two partons and LO-accurate MEs for up to four partons.
Samples of diboson final states ($VV$) were also simulated using the \SHERPA~{2.2.1} generator.
The production of SM \tttt\ events is modeled using the \MGNLO~{2.6.2} generator at NLO with
the \texttt{NNPDF3.1NLO}~\cite{Ball:2014uwa} PDF.
The rare process of single-top-quark associated \(Z\) boson production, \(tZ\), is modeled using the \MGNLO generator at LO.
 
All events generated using \POWHEGBOX or \MGNLO are interfaced with \PYTHIA{8.230}~\cite{Sjostrand:2014zea} using the \textsc{A14}
tune~\cite{ATL-PHYS-PUB-2014-021} and the \texttt{NNPDF2.3LO}~\cite{Ball:2014uwa} PDF set. Events generated using \SHERPA employ the set of tuned parameters developed by the \SHERPA authors.
 
To assess the uncertainty due to the choice of generator, the \ttbar\ and single-top-quark
samples produced with the nominal generator set-ups are
compared with alternative samples generated with \MGNLO
(interfaced with \PYTHIA{8}) for the calculation of the hard-scattering processes.
These alternative samples were generated using the same PDF in the ME as in the nominal samples.
Additional \ttbar\ and single-top-quark samples were produced
by replacing \PYTHIA{8} with \HERWIG~{7.04}~\cite{Bahr:2008pv,Bellm:2015jjp}
for parton showering and hadronisation, using the H7UE
tune~\cite{Bellm:2015jjp} and the
\texttt{MMHT2014LO}~\cite{Harland-Lang:2014zoa} PDF set.
These samples are used to evaluate uncertainties due to the choice
of parton shower and hadronisation model.

% End of text imported from the .//samples/samples.tex input file

% The next lines are included from the .//analysis/analysis.tex input file
\section{Analysis strategy}
\label{sec:strategy}
 
This analysis targets events where both resonance top quarks decay hadronically.
For a heavy resonance signal with $\mZp \ge 1.0~\tev$,
the two resonance top quarks are expected to be highly boosted.
Large-$R$ jets are used as proxies for the two highly boosted top quarks.
The invariant mass distribution of the two large-$R$ jets with the highest \pt,
labelled as \mJJ,
is scanned for an excess over the background prediction.
 
\begin{figure}[t]
\centering
\subfloat[]{
\includegraphics[width=0.47\textwidth]{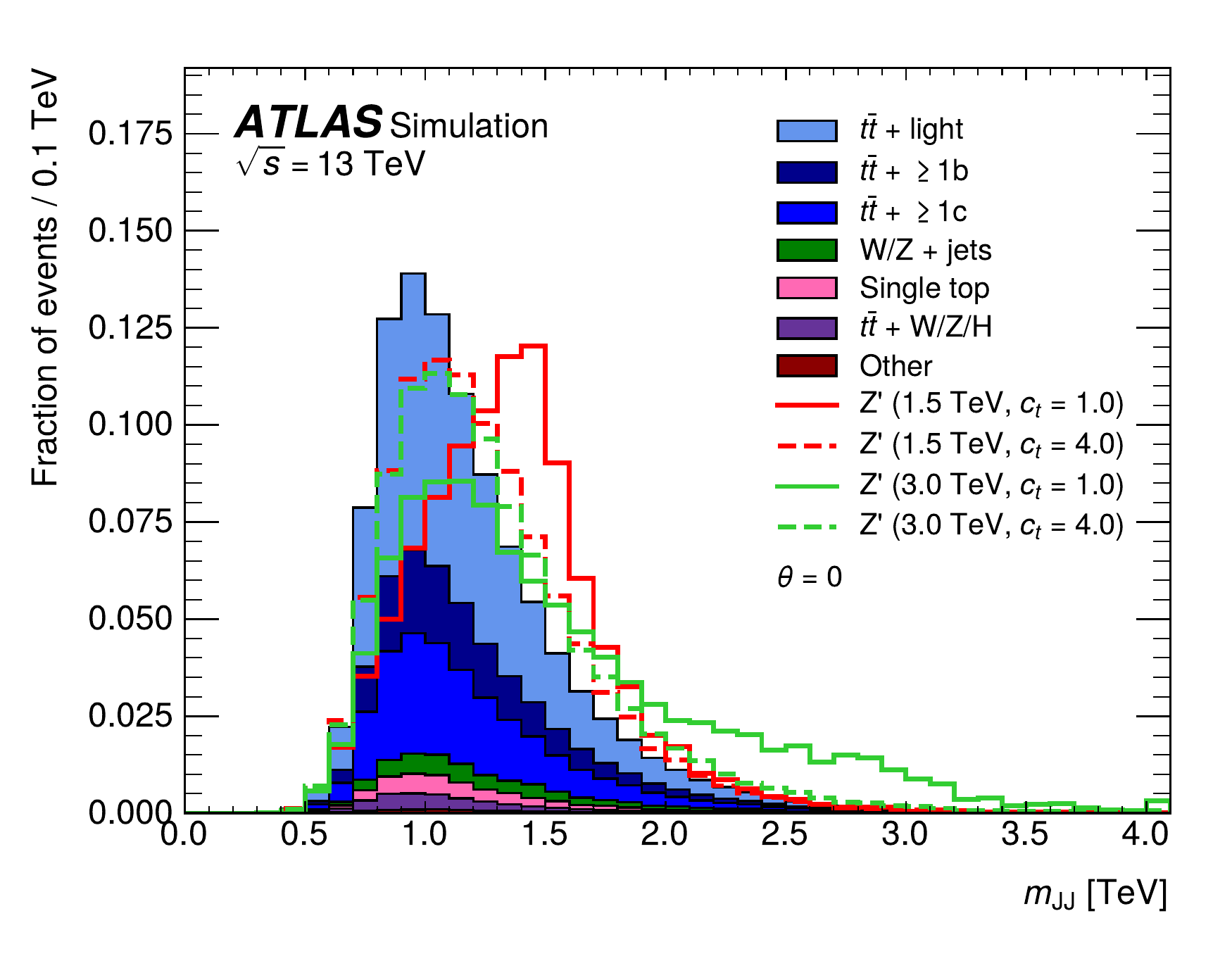}
}
\subfloat[]{
\includegraphics[width=0.47\textwidth]{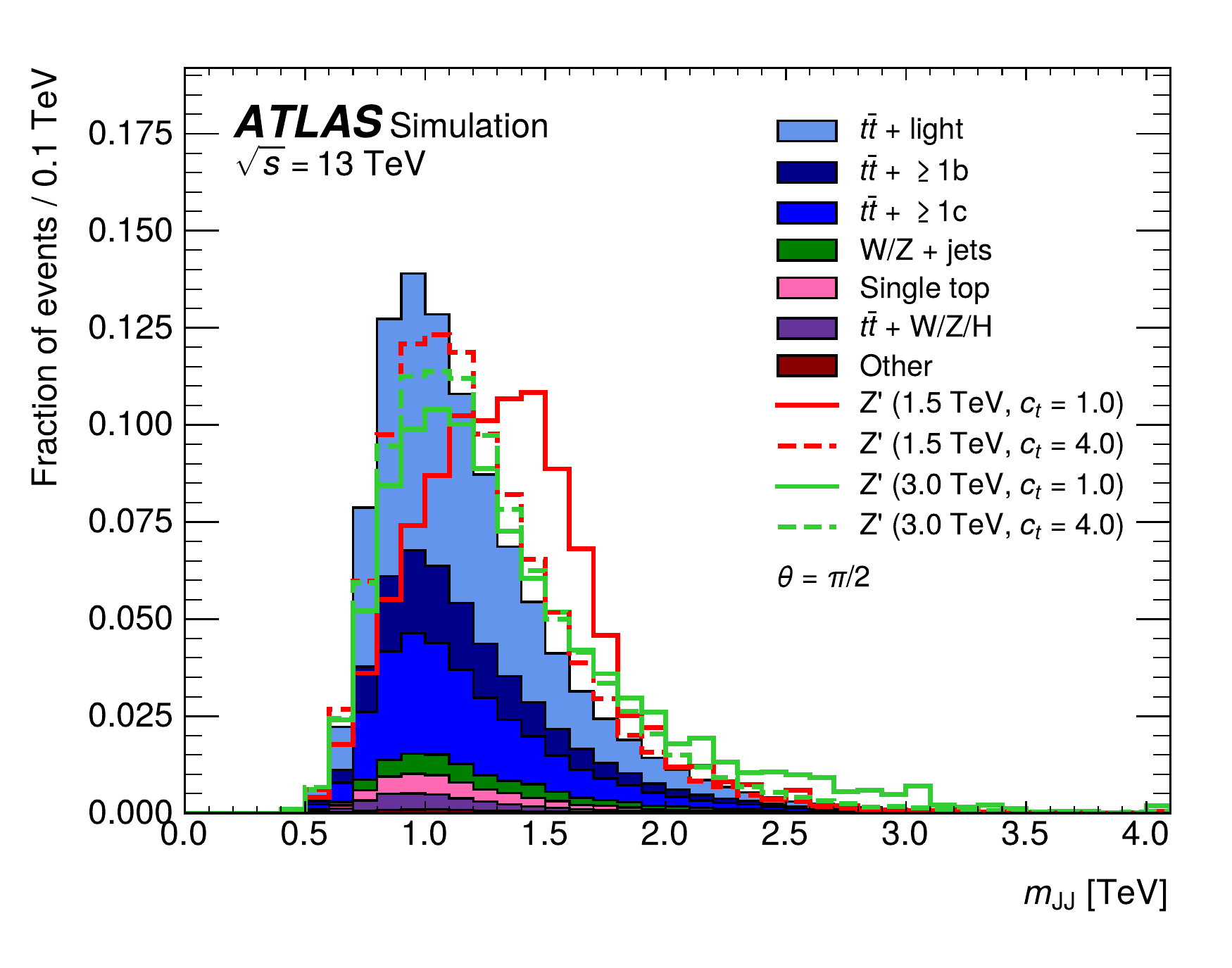}
}
\caption{
The \mJJ\ distributions from the top-philic $Z'$ signals compared to the total background after the event preselection.
The signal distributions are shown for two values of the chirality angle, (a) $\theta=0$ and (b) $\theta=\pi/2$.
In each case, four signal samples with different mass resolutions are presented for the possible combinations of $\mZp = 1.5$~\TeV, $3$~\TeV\ and $c_t=1$, $4$.
All predictions are given by MC simulations.
The contributions from the SM \tttt, diboson, and $VH$ processes are combined in the category `Other'.
The total background and all signal distributions are normalised to unit area.
}
\label{fig:signal_vs_background}
\end{figure}

The \mJJ\ distributions for the top-philic $Z'$ signal with different \mZp\ and $c_t$ values are shown in Figure~\ref{fig:signal_vs_background}.
A peak near the generated value
of \mZp\ is typically observed for masses \mZp\ $\leq 2$ TeV and coupling strengths $c_t\leq 1$.
The peak at the resonance mass in general is wider for higher $Z'$ masses and for higher $c_t$ values due to the increase of the $Z'$ natural width.
All distributions are skewed towards the lower mass end for multiple reasons.
One factor is the contamination from large-$R$ jets
that do not capture or only capture part of the decay products from a hadronically-decaying resonance top quark.
Moreover, the parton luminosity effect and QCD radiation from the highly boosted top quarks
also contribute to the lower mass side.
The difference between the two chirality hypotheses, $\theta=0$ and $\theta=\pi/2$, is small,
with a slightly better mass resolution in case of $\theta=0$.
This is due to the maximal contribution from $tjZ'$ and $tWZ'$ production modes,
which have smaller ambiguity in reconstructing the $Z'$ resonance using the two large-$R$ jets given the lower expected jet and b-jet multiplicities.

\begin{figure}[t]
\centering
\subfloat[]{
\label{fig:analysis:nadd}
\includegraphics[width=0.47\textwidth]{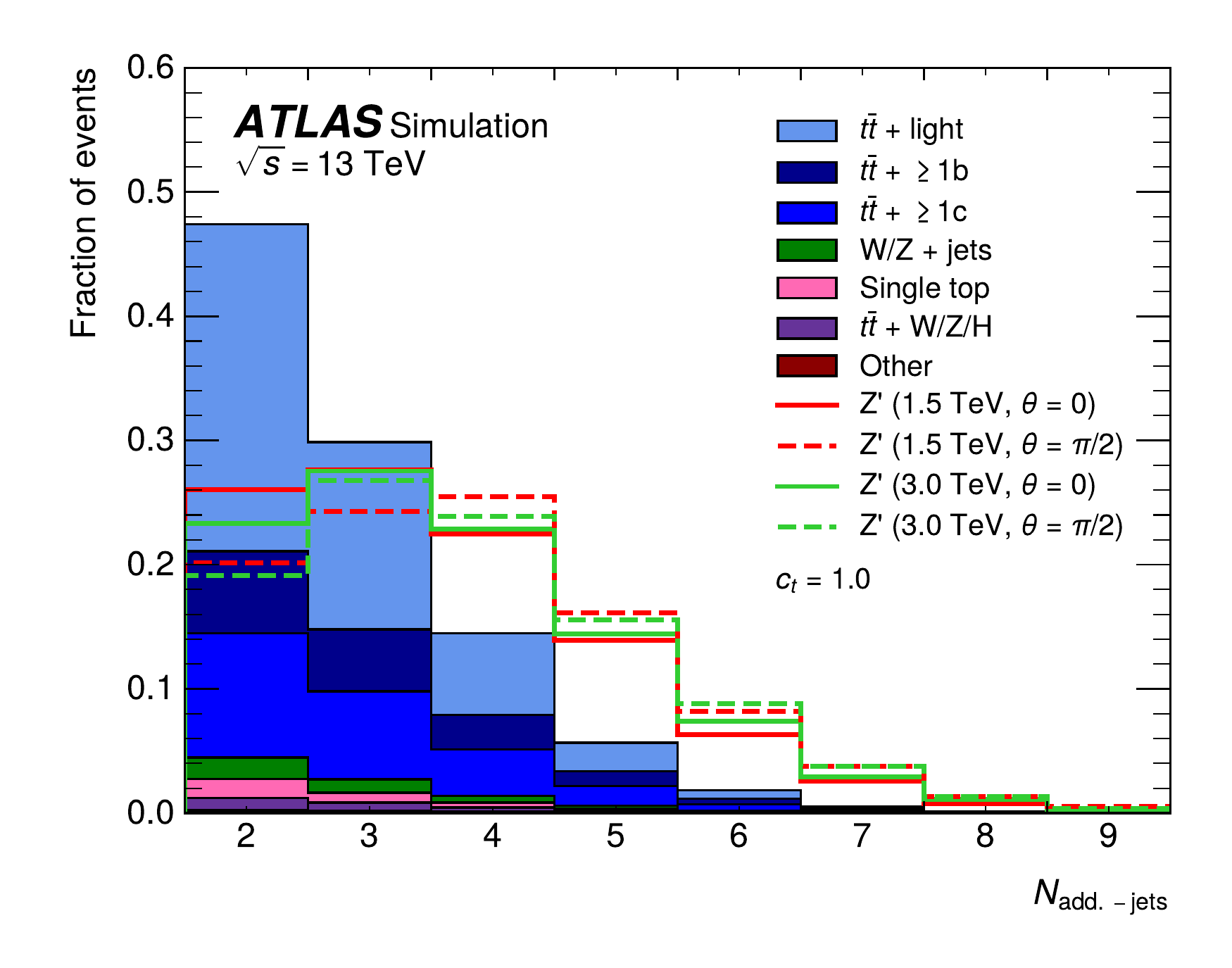}
}
\subfloat[]{
\label{fig:analysis:nb}
\includegraphics[width=0.47\textwidth]{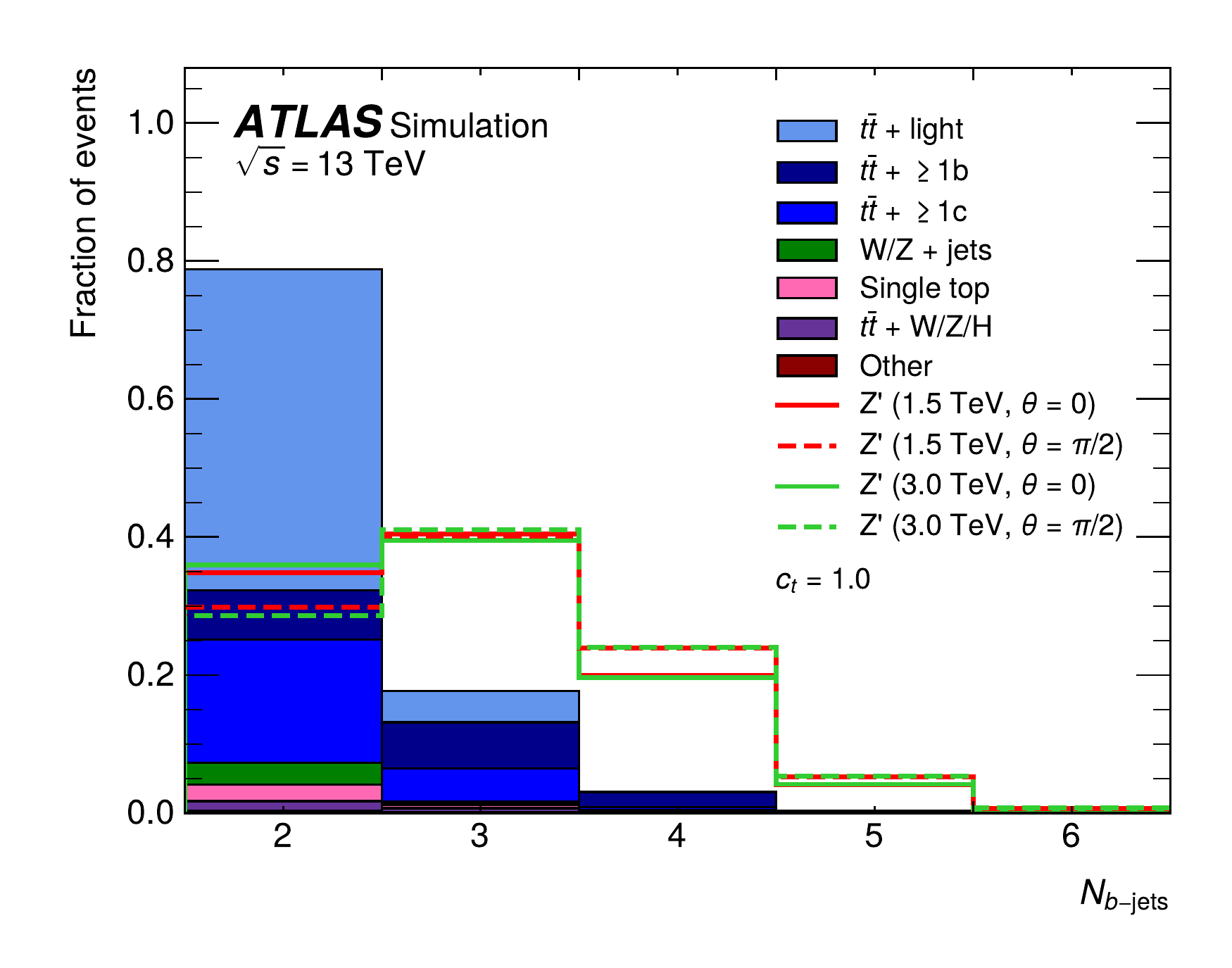}
}
 
\caption{
The (a) \naddjets\ and (b) \nbjets\ distributions from the top-philic $Z'$ signals compared to the total background after the event preselection.
The predictions are given by MC simulations.
The contributions from the SM \tttt, diboson, and $VH$ processes are combined in the category `Other'.
Four signal samples are presented with $\mZp = 1.5$ and 3.0~\TeV\ and $\theta = 0$ and $\pi/2$. 
All distributions are normalised to unit area.
}
\label{fig:nadd_nb}
\end{figure}

Despite the above effects that degrade the resolution for reconstructing the $Z'$ mass,
the \mJJ\ distributions for the signal with small \mZp\ and $c_t$ are still distinct from those for background, as shown in Figure~\ref{fig:signal_vs_background}.
For the signal hypotheses with larger \mZp\ and $c_t$, the distributions become more background-like given the increasing resonance width and the low-mass shoulder.
According to MC simulations, about 90\% of background after the preselection consists of \ttjets\ events.
Other background events mainly come from the production of \ttZ, \ttW, \ttH, single top-quark and $W/Z$ bosons in association with jets.
The SM \tttt, diboson and other rare processes contribute less than 1\% of all background events.
For background events, a smoothly falling \mJJ\ distribution is expected once \mJJ\ passes the turn-on threshold
due to the minimum requirements on the large-$R$ jet mass and \pt.
 
The search is performed in the range of the \mJJ\ distribution between 1.0~\TeV\ and 3.2~\TeV.
The lower bound was chosen to minimise the impact due to the turn-on threshold on the background modelling,
whereas the upper bound was motivated by the vanishing amount of expected SM background events at the high end of the \mJJ\ spectrum.
The product of the efficiency and acceptance of the signal  in the \ttZp\ production mode after the preselection varies from 3.1\% to 7.2\% for the different signal hypotheses, estimated using the simulated samples described in Section~\ref{sec:samples}.

\begin{figure}[!h]
\centering
\includegraphics[width=0.5\textwidth]{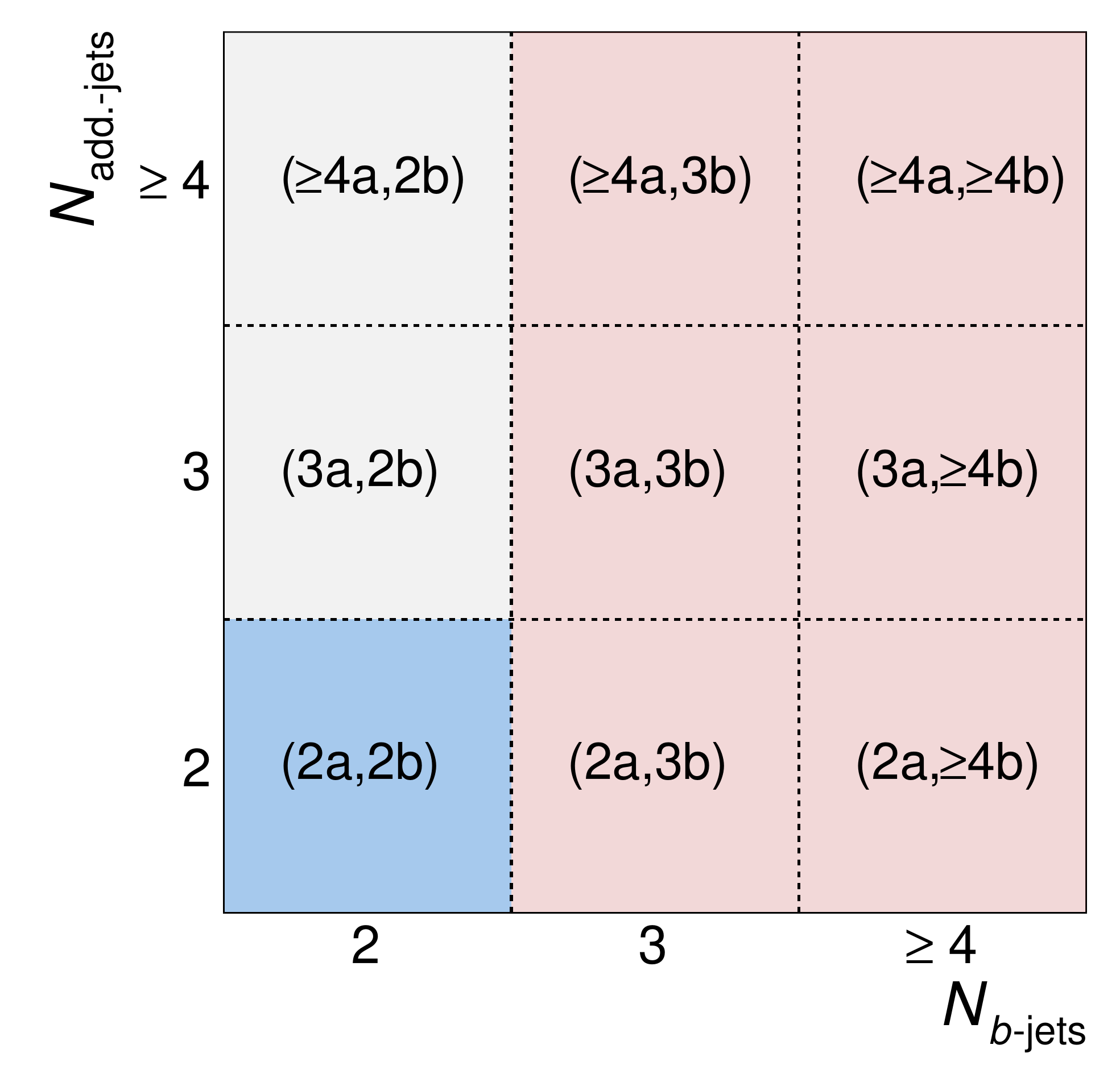}
\caption{
An illustration of the analysis regions defined using \naddjets\ and \nbjets.
The source, validation and signal regions are highlighted in blue, gray and red, respectively.
The names of the regions used in the rest of the document are also shown.
}
\label{fig:regions}
\end{figure}

In the single-lepton final state, $Z'$ signal events are expected to have a high multiplicity of jets in addition to the two large-$R$ jets (\naddjets ) and the $b$-tagged jets (\nbjets ).
The distributions of \naddjets\ and \nbjets\ are shown for the signals and the background in Figure~\ref{fig:nadd_nb},
demonstrating clear discriminating power to separate these contributions.
These two variables are used to categorise events into regions with different signal-to-background ratios.
As illustrated in Figure~\ref{fig:regions},
nine regions are defined, denoted by `N$a$', `M$b$',
where N$a$ and M$b$ represent \naddjets\ and \nbjets, respectively, and both range from 2 to ${\ge}4$.
The region with the lowest signal contamination, \twoatwob,
is referred to as the `source region' and used to derive the data-driven background estimate
detailed in Section~\ref{sec:background}.
The highest \naddjets\ and \nbjets\ regions are chosen according to the expected signature of the \ttZp\ events and help to maintain a statistically significant background estimation.
All regions with at least three $b$-tagged jets have relatively higher expected signal contributions compared to other regions.
They are used for signal extraction and are referred to as the `signal regions'.
In the \threeatwob\ and \fouratwob\ regions,
the small signal contribution provides little gain in sensitivity,
and the expected signal contamination is not negligible for the data-driven background estimate.
Therefore they are not used as either signal nor source region.
However, they are used as validation regions to verify the modelling of the background.
Scaling the top-philic simplified signal model to the expected exclusions achieved by this search, the largest signal contamination is below 2\% in the source region and below 5\% in the validation regions, depending on the model parameter choices.
Figure~\ref{fig:nadd_nb} illustrates the background composition
in bins of \naddjets\ and \nbjets\ expected from the MC predictions.
The fractions of non-\ttbar\ background are similar in most bins.
The most dramatic change is in the fraction of \ttbjets\ events.
From MC simulations, the fraction of \ttbjets\ events with respect to the total sum of background events is expected to increase from 8\% in regions with two $b$-tagged jets to 68\% in the most signal-like regions with at least four $b$-tagged jets.

% End of text imported from the .//analysis/analysis.tex input file

% The next lines are included from the .//prefitBack/prefitBack.tex input file
\section{Background estimation}
\label{sec:background}
 
The search for the resonance in the \mJJ\ distribution
heavily relies on the modelling of the continuum background.
However, the MC prediction is expected to be unreliable
and susceptible to large modelling systematic uncertainties
due to various factors discussed in the following.
The top-philic $Z'$ signal in the one-lepton channel features a large jet multiplicity.
For the main background, \ttjets, events with a similar final state contain multiple partons beyond Born level.
This means that in the nominal \ttjets\ sample generated using NLO ME,
most of the jets in the final state are from the parton shower with limited precision~\cite{TOPQ-2018-15}.
These jets form the large-$R$ jets used to construct \mJJ.
Additional mismodelling and uncertainties come from the prediction of the boosted large-$R$ jets,
which are subject to the often badly modelled collinear radiation.
Finally, the \ttHF\ background that populates the signal regions
is underestimated by the current MC predictions~\cite{TOPQ-2017-12,CMS-TOP-18-011}.

A data-driven technique is used to provide a reliable background prediction.
It is based on the similarity of the background \mJJ\ shape in all analysis regions observed in MC simulations.
Therefore, the shape of the background is obtained from data in the source region where there is negligible signal contamination.
The background shape is then extrapolated to the signal regions.
The extrapolation factors are derived from the MC simulations,
accounting for the reduced event rate in higher \naddjets\ and \nbjets\ bins
and the minor differences in the shape of \mJJ\ between the regions.
The final background model is obtained using a profile likelihood fit
via further adjustments according to all systematic variations considered in this analysis.

Firstly, the shape of the background is obtained by fitting the data in the source region, \twoatwob,
using the following parameterization~\cite{EXOT-2019-03,EXOT-2018-48}:
\begin{equation}
f(x) = (1-x)^{p_1} \cdot x^{p_2 + p_3 \cdot \log{(x)}},
\label{eq:dijet}
\end{equation}
where $x = \mJJ/\sqrt{s}$, with $\sqrt{s}$ being the $pp$ collision center-of-mass energy,
and $p_1$ to $p_3$ are the fit parameters controlling the shape of the function.
The fit result is presented in Figure~\ref{fig:dijet_fit},
showing that the fitted function describes the data well within the uncertainties due to the limited number of data events.
Alternative parameterizations considering additional parameters or alternative functional forms have been tested,
but were found to provide no improvement in the performance of the background estimate.

\begin{figure}[!h]
\centering
\includegraphics[width=0.7\textwidth]{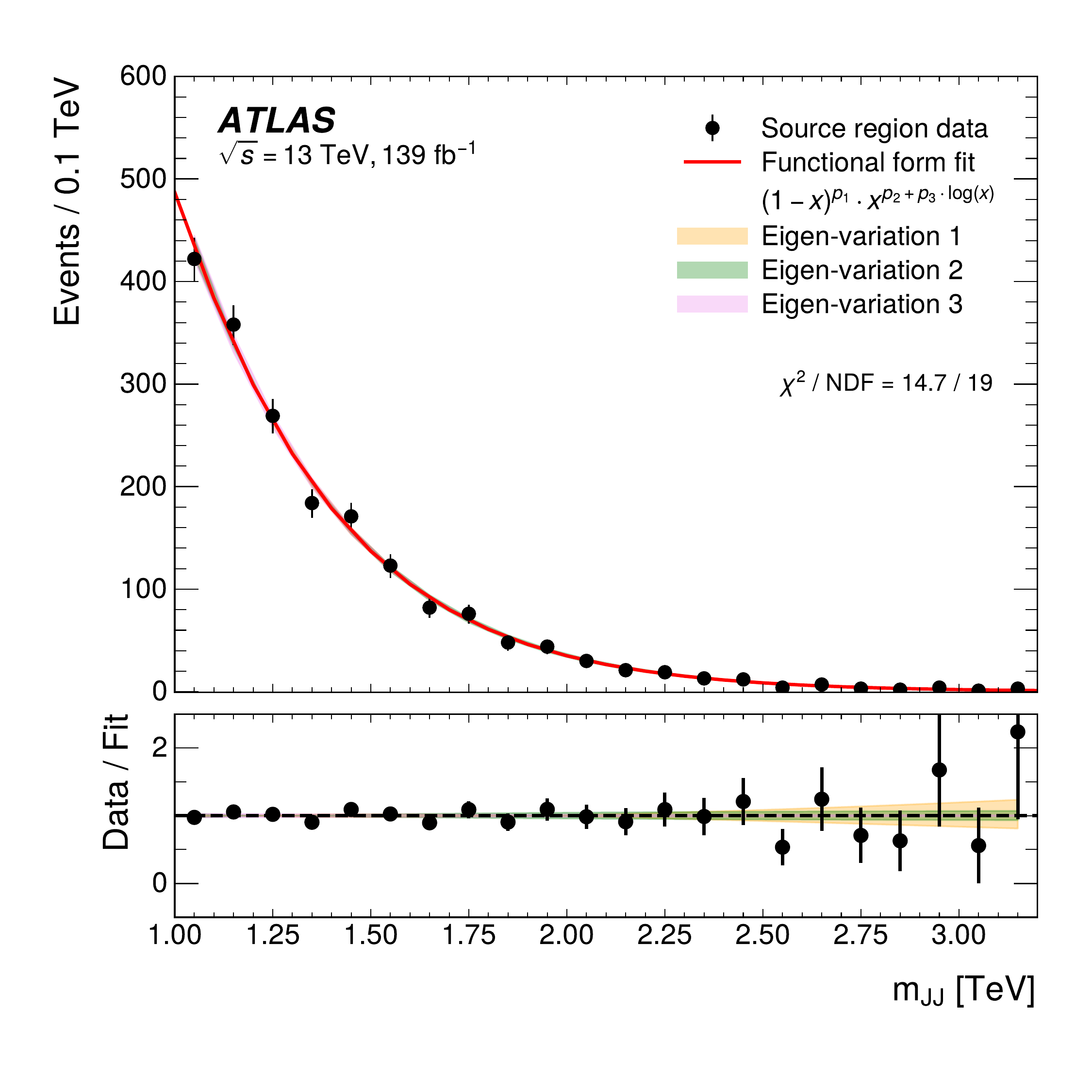}
\caption{
The distribution of data in the source region, along with the fitted function to data.
The bands show the three independent variations obtained by diagonalising the covariance matrix of the fit.
The third eigen variation is not visible due to its smallness.
}
\label{fig:dijet_fit}
\end{figure}

The fitted function from the previous step is scaled by extrapolation factors for each signal region,
in order to obtain the background prediction in that region.
The extrapolation factors are derived from MC simulation in bins of \mJJ,
as the ratio of the expected number of events in the region of interest
to that in the source region.
To minimize the impact due to the limited number of MC events,
the function in Equation~\ref{eq:dijet} is used, fitting the \mJJ\ distribution
predicted by MC simulation in both the region of interest and the source region.
The extrapolation factors are then obtained by taking the ratio of these fitted functions.
The shape of the background in the various analysis regions differs by up to $\pm15\%$
with respect to that in the source region,
according to the prediction of the extrapolation factors.
The extrapolation is performed with all backgrounds combined.
The shape of the \mJJ\ distribution from the different background processes are compared using MC simulations.
For the \ttjets, single top-quark and $t\bar{t}Z/W/H$ processes, constituting about 95\% of the total background,
the shapes are consistent within the uncertainties due to the limited number of MC events.
Furthermore, the composition and the shape of the small backgrounds, i.e. those other than the \ttjets backgrounds,
are found to be similar across all analysis regions.
As an additional check, alternative extrapolation factors were derived with \ttjets\ MC simulation only,
to account for the effect due to variation in the non-\ttbar\ backgrounds.
The resulting difference was found to be smaller than the uncertainty due to the limited number of MC events, and is therefore neglected.
 
The extrapolation factors are based on the MC predictions in each of the analysis regions.
Systematic uncertainties affecting the predicted distributions (\mJJ, \naddjets\ and \nbjets) are therefore
propagated to the background estimate via the extrapolation factors.
These include modelling uncertainties in all background MC samples and experimental uncertainties.
For each systematic variation, the extrapolation factors are rederived using the same procedure as for the nominal background prediction.
To account for the deficit of \ttHF\ events in MC simulation,
additional normalisation uncertainties are assigned to \ttcjets\ and \ttbjets\ events.
Furthermore, dedicated uncertainties in the data-driven background estimate and the extrapolation are considered.
The details on the systematic uncertainties included in this analysis are described in Section~\ref{sec:sys}.
All systematic uncertainties are incorporated in a profile likelihood fit to data in the signal regions,
as discussed in Section~\ref{sec:statAna}.

% End of text imported from the .//prefitBack/prefitBack.tex input file

% The next lines are included from the .//sys/sys.tex input file
\section{Systematic uncertainties}
\label{sec:sys}
 
Different sources of systematic uncertainty affect the search presented here,
including those related to the luminosity, the identification and reconstruction of the physics objects,
the MC simulation of the signal and background processes,
and the method used to estimate the background.
In the following, a brief description of the sources of systematic uncertainty is provided.
A particular emphasis is put on the ones related to the \ttbar background prediction,
which will be shown to have the largest impact on the sensitivity of the search.
The systematic variations can affect the normalisation of the signal and background templates
estimated in the different regions as well as the shape of the \mJJ\ distributions.
All systematic uncertainties on the background prediction enter the analysis exclusively via
extrapolation factors, except for the uncertainties related to the functional fits to data and MC described in Section~\ref{subsec:dijetUnc},
and the signal bias uncertainty described in Section~\ref{subsec:SignalBias}.
The luminosity uncertainty only applies to the signal.

% The next lines are included from the .//sys/experimental.tex input file
\subsection{Experimental uncertainties}
\label{subsec:expUnc}

The uncertainty in the combined 2015--2018 integrated luminosity is 1.7 \% \cite{ATLAS-CONF-2019-021}, obtained using the LUCID-2 detector \cite{LUCID2} for the primary luminosity measurements.
This uncertainty affects the signal prediction in the model-dependent interpretation.
 
Other experimental uncertainties arise from corrections and calibrations applied to MC simulations.
These uncertainties affect the background estimate via the extrapolation factors,
as well as the MC prediction of the $Z'$ signals.
An uncertainty is considered for the reweighting factors that correct the pile-up profile in MC simulations to match those in data.
Uncertainties on the modelling of leptons arise from their momentum and energy scale calibration and resolution,
as well as the trigger, reconstruction, identification, isolation efficiencies~\cite{EGAM-2018-01,PERF-2015-10}.
Uncertainties on the modelling of jets mainly come from their energy scale (JES) and resolution (JER),
containing effects from jet flavour composition, single-particle response, and pile-up~\cite{ATL-PHYS-PUB-2015-015,PERF-2011-03}.
An uncertainty is assigned for the efficiencies of the JVT requirement on jets~\cite{PERF-2014-03}.
Uncertainties are considered for the calibration of the $b$-tagging efficiencies,
including the efficiencies of tagging $b$-jets as well as the rates of mis-tagging $c$-jets and light-flavour jets~\cite{FTAG-2018-01,FTAG-2020-08,FTAG-2019-02}.

% End of text imported from the .//sys/experimental.tex input file

% The next lines are included from the .//sys/dijet.tex input file
\subsection{Uncertainties on functional fit and extrapolation}
\label{subsec:dijetUnc}
 
Dedicated uncertainties are assigned for the data-driven background estimate
and the MC extrapolation from the source region to other regions.
The uncertainties from each functional fit
are obtained from the covariance error matrix of the fit.
Three statistically independent variations are extracted from an eigen-decomposition of the covariance matrix.
This leads to three uncertainties associated to
the functional fit in the source regions to data and MC, respectively. In addition, three uncertainties are also associated
to the fit of the MC in each of the six signal regions which leads to 24 components of uncertainty in total.
The resulting uncertainties due to the functional fit to data and MC in the source region are both correlated
across the signal regions.

% End of text imported from the .//sys/dijet.tex input file

% The next lines are included from the .//sys/ttbarModel.tex input file
\subsection{Theoretical uncertainties for the \ttbar background}
\label{subsec:ttbarUnc}
 
An uncertainty of $50\%$ in the normalisation of the \ttbjets\ events as well as the \ttcjets\ events is applied~\cite{TOPQ-2017-12,CMS-TOP-18-011},
and an uncertainty of $10\%$ is considered for the \ttljets\ events.
The uncertainties due to the choice of generator and parton shower model used to simulate
the inclusive \ttbar sample are evaluated by comparing the nominal \ttbar sample with
alternative \ttbar samples, detailed in Section~\ref{sec:samples}.
The uncertainties associated with the choice of NLO generator are estimated by comparing the predictions of \POWHEGBOX and \mgamc.
The effect of the choice of parton shower and hadronisation model is estimated from comparing the prediction of
\POWHEG+ \PYTHIA~{8} to \POWHEG+ \HERWIG~{7.04}.
These two uncertainties are split into four components, each affecting either only the shape of the \mJJ\ distributions or the acceptance and migration for the $3b$ regions and the ${\ge}4b$ regions separately.
This treatment is motivated by the different compositions of \ttljets, \ttcjets\ and \ttbjets\ events in the two $b$-jet multiplicity bins and the large effect on the overall acceptance and migration of the events across the regions due to different parton shower and hadronisation models.
Uncertainties due to missing higher-order QCD corrections are estimated by separately
varying the renormalisation and the factorisation scales by factors of \num{2.0} and \num{0.5}
in the nominal \ttbar sample and taking the envelope.
Additionally, uncertainties in the amounts of initial- and final-state radiation (ISR and FSR)
from the parton shower (PS) are assessed by respectively varying the corresponding
parameter of the A14 PS tune\footnote{The A14 tune's variations
of $\alpha_\mathrm{S}^\mathrm{ISR}(m_Z)$ are usually referred to as Var3c, and correspond
to $\alpha_\mathrm{S}^\mathrm{ISR}(m_Z)=0.127^{+0.013}_{-0.012}$~\cite{ATL-PHYS-PUB-2014-021}.}
and by varying the FSR renormalisation scale by factors of \num{2.0} and \num{0.625}.
The uncertainty related to the parton distribution function (PDF) is evaluated by using the
PDF4LHC systematic variations~\cite{Butterworth:2015oua}.

% End of text imported from the .//sys/ttbarModel.tex input file

% The next lines are included from the .//sys/nonttbar.tex input file
\subsection{Modelling uncertainties for non-\ttbar backgrounds}
\label{subsec:nonttbarUnc}
 
Non-\ttbar background processes represent a minor fraction of the total background.
Therefore, the relevant uncertainties have a small impact on the result of the analysis.
Similar to the \ttbar background, uncertainties due to missing higher-order QCD corrections,
the amounts of initial- and final-state radiation, and the parton distribution function
are also considered for most of the non-\ttbar backgrounds.
 
An uncertainty of 30\% in the total cross section of the three
single-top-quark production modes is included.
This conservative number is motivated by the uncertainties
due to the modelling of several associated jets and heavy flavour jets in the production.
Uncertainties associated with the parton shower and hadronisation model and the generator choice
are evaluated by comparing the nominal \POWHEG+~\PYTHIA~{8} sample for each process with
alternative samples produced with \POWHEG+~\HERWIG~{7} and \AMCatNLO+~\PYTHIA~{8}.
An additional uncertainty on the interference between \tW and \ttbar
production at NLO is evaluated by comparing the
nominal $tW$ sample produced using the diagram removal scheme with an alternative
sample produced with the same generator but using the `diagram subtraction' scheme~\cite{Frixione:2008yi}.
 
Modelling uncertainties in the \ttW, \ttZ, and \ttH processes are evaluated in a
similar way. Uncertainties of $60\%, 15\%,$ and $20\%$ are applied
to the \ttW, \ttZ, and \ttH cross sections, respectively~\cite{deFlorian:2016spz,HIGG-2018-57,Aad:2020klt}.
 
An uncertainty of 60\% is assumed for the \Vjets production cross section.
It is estimated by adding a 24\% uncertainty in quadrature for each additional jet
based on a comparison among different algorithms for merging LO matrix elements
and parton showers~\cite{Alwall:2008epjc}. An uncertainty of 20\% is assumed for the SM \tttt\
production cross section~\cite{ATLAS:2021kqb}. A conservative cross section uncertainty of 50\%
is applied to all other small background processes.

% End of text imported from the .//sys/nonttbar.tex input file

% The next lines are included from the .//sys/signalBias.tex input file
\subsection{Signal bias uncertainty}
\label{subsec:SignalBias}
 
A dedicated uncertainty is considered in the model-dependent interpretation of the results,
accounting for the bias on the extraction of the signal yields caused by the background model.
The details on the signal extraction and the model-dependent interpretation are described in Section~\ref{sec:statAna}.
To evaluate the size of the bias, a large number of pseudo-datasets are sampled using the expected background directly from MC simulations.
The model-dependent signal extraction (as described in Section~\ref{subsec:model_dep_interp}) is performed for each of these pseudo-datasets,
taking into account all systematic uncertainties described previously.
The resulting distribution of the signal strengths is fitted with a Gaussian function.
The deviation of the fitted central value from zero is taken as the size of the bias.
This procedure is repeated for all signal mass points.
The three points with the largest bias are used to compute a second order polynomial function.
The eventual size of the bias for each mass point is evaluated from the fitted polynomial function. To obtain dedicated uncertainties for all choices of model parameters, this procedure is repeated for all distinct $c_t$ and chirality parameter $\theta$ values explored in this search.
The uncertainty is determined using the MC signal template scaled by the corresponding value of bias,
and is implemented as an uncertainty on the background expectation. Depending on the analysis region and signal point, this translates to uncertainties up to 14\% in the individual signal regions, with the majority of values being a few percent in size.

% End of text imported from the .//sys/signalBias.tex input file

% End of text imported from the .//sys/sys.tex input file

% The next lines are included from the .//statistical/statistical.tex input file
\section{Statistical analysis}
\label{sec:statAna}
 
The search for a resonance signal is conducted using the binned \mJJ\ distributions in all signal regions.
An approach with minimal model dependence is adopted, followed by a model-dependent interpretation.
A profile likelihood fit is used to obtain the final background model
and make further statistical inference regarding the presence of a signal.
The search with minimal model dependence is performed using \textsc{BumpHunter}~\cite{Choudalakis:2011qn},
a hypothesis-testing tool that searches for local data excesses compared to the expected background.
To obtain the expected background, the profile likelihood fit is performed with a background-only hypothesis.
In the model-dependent interpretation, the $Z'$ signal samples described in Section~\ref{sec:samples}
are used to interpret the data.
Profile likelihood fits are performed with the signal-plus-background hypothesis,
testing the compatibility between data and the models with different \mZp , $c_t$ and chirality parameter $\theta$.

\subsection{Profile likelihood fit}
 
The statistical analysis is based on a binned profile likelihood function.
The bin width is chosen to be 100~\GeV,
with two large bins of 500~\GeV\ and 700~\GeV\ at the high mass end of the \mJJ\ spectrum to avoid empty bins.
Each bin in the signal regions is represented by a Poisson probability term for the observed data,
with the total expected yield given by the background model described in Section~\ref{sec:background}
and the $Z'$ signal samples.
The likelihood function $\mathcal{L}(\mu, \pmb{\theta})$
is constructed as the product of the Poisson probability terms over all bins.
This function depends on the signal-strength parameter $\mu$,
a multiplicative factor to the assumed pre-fit signal cross section,
and $\pmb{\theta}$, a set of nuisance parameters encoding the effect of systematic uncertainties.
The nuisance parameters are implemented in the likelihood function as Gaussian or log-normal constraints.
The total number of expected events in a given bin therefore depends on $\mu$ and $\pmb{\theta}$.
The fit is performed by finding the values of $\mu$ and $\pmb{\theta}$ that maximise the likelihood function.
In the case of a fit with the background-only hypothesis $\mu$ is fixed to 0.
The nuisance parameters $\pmb{\theta}$ allow variations of the expectations for signal and background
according to the corresponding systematic uncertainties,
whilst penalising the likelihood function for any deviation according to the specified constraints.
The fit also reduces the impact of systematic uncertainties on the search sensitivity
by exploiting the highly populated and background-dominated regions.
The result of the profile likelihood fit is further used for the signal extraction.

\subsection{Search with minimal model dependence}
 
In the search using \textsc{BumpHunter}, the data and the expected background in each of the six fitted regions
are compared in sliding windows of variable sizes.
The smallest window is required to contain two bins with the binning shown in \Figs{\ref{fig:postfit_BH_1}}{\ref{fig:postfit_BH_2}} in Section~\ref{sec:results},
corresponding to a width that is slightly smaller than the expected resolution extracted from the $Z'$ MC samples.
The Poisson probability is evaluated for all windows.
For each region, the window with the smallest Poisson probability is chosen to be the most prominent window.
The corresponding probability $p_{\text{min}}$ is used to construct
the \textsc{BumpHunter} test statistic $t$ defined as
\begin{align}
t=\left\{
\begin{array}{ll}
0 & \text{if } d \leq b \\
-\log{p_{\text{min}}} & \text{otherwise}
\end{array}
\right.,
\label{Eq:BumpHunterTestStat}
\end{align}
where $d$ and $b$ represent the number of observed data and the expected background events in the window, respectively.
A deviation of data from background is only considered as evidence against the background-only hypothesis if data exceed the expectation.
To obtain the local $p$-value and the significance of the most interesting bump,
$10^5$ pseudo-experiments are sampled from the expected background,
and the $t$ value from data is compared to the distribution of the $t$ values from the pseudo-experiments.
The global $p$-value and significance are then computed for the most prominent window of each region,
taking into account the trial factors that incorporate the look-elsewhere effect~\cite{Gross_2010}.

\subsection{Model-dependent interpretation}
\label{subsec:model_dep_interp}
 
The model-dependent interpretation is based on the signal samples described in Section~\ref{sec:samples}.
The signal strength is extracted by performing the profile likelihood fit with the signal-plus-background hypothesis.
The fit and the subsequent statistical analysis are performed for each mass point.
The test statistics are defined based on the likelihood ratio
$\lambda(\mu) = {\mathcal L}(\mu,\pmb{\hat{\hat{\theta}}}_\mu)/{\mathcal L}(\hat{\mu},\pmb{\hat{\theta}})$,
where $\hat{\mu}$ and $\pmb{\hat{\theta}}$ are the values of the parameters that maximise the likelihood function,
and $\pmb{\hat{\hat{\theta}}}_\mu$ are the values of the nuisance parameters
that maximise the likelihood function for a given value of $\mu$.
To evaluate the compatibility of the observed data with the background-only hypothesis,
the test statistic $q_0$, defined as
\begin{align}
q_0=\left\{
\begin{array}{ll}
-2\ln\lambda(0) & \text{if } \hat\mu \geq 0 \\
0 & \text{otherwise}
\end{array}
\right.,
\label{Eq:q_0}
\end{align}
is used.
The resulting $p$-value represents the compatibility $p_0$.
A small $p_0$ indicates that the background-only model does not describe the data well,
and the significance of the corresponding signal is further examined.
In the absence of any significant excess above the background expectation,
upper limits on the signal production cross section at 95\% confidence level are derived.
The test statistic $q_\mu$, defined as
\begin{align}
q_\mu=\left\{
\begin{array}{ll}
-2\ln\lambda(\mu) & \text{if } \hat\mu \leq \mu \\
0 & \text{otherwise}
\end{array}
\right.,
\label{Eq:q_mu}
\end{align}
is used with the $\text{CL}_{\textrm{S}}$ method~\cite{Junk_1999,Read:2002hq}.
Both $p_0$ and the upper limits are computed using the asymptotic approximation~\cite{Cowan:2010js}.
All statistical analyses are performed using the \textsc{RooStats} framework~\cite{Cranmer:1456844,RooFit,moneta2011roostats}.

% End of text imported from the .//statistical/statistical.tex input file

% The next lines are included from the .//results/results.tex input file
\section{Results}
\label{sec:results}
 
The results of the statistical analysis are presented in this section.
Prior to the analysis using real data,
the profile likelihood fit model and the statistical methods were tested against pseudo-datasets constructed using simulated events.
Both the background modelling and the signal extraction were validated,
using background-only and signal-plus-background pseudo-datasets.
In particular, alternative \ttbar\ background predictions were considered when building the pseudo-datasets.
They were used to stress-test the fit, especially the incorporated systematic uncertainties related to \ttbar\ background modelling.
These alternative predictions include an enhanced composition of the \ttHF\ backgrounds,
as well as a \ttbar\ sample generated with \SHERPA{2.2.10}\footnote{
This \ttbar\ MC sample was generated with \SHERPA{2.2.10} using NLO-accurate matrix elements for up to one additional parton,
and LO-accurate matrix elements for up to four additional partons.
The additional parton emissions are matched and merged with the \SHERPA parton shower
based on Catani-Seymour dipole factorisation~\cite{Gleisberg:2008fv,Schumann:2007mg}
using the MEPS@NLO prescription~\cite{Hoeche:2011fd,Hoeche:2012yf,Catani:2001cc,Hoeche:2009rj}.
},
which is not used in the definition of any uncertainty.
These stress-tests demonstrated the capability of the fit to constrain the relevant systematic uncertainties,
and to model the data correctly in case of mismodelling in the extrapolation factors from MC predictions.
 
\begin{figure}[t]
\centering
\subfloat[]{\includegraphics[width=0.495\textwidth]{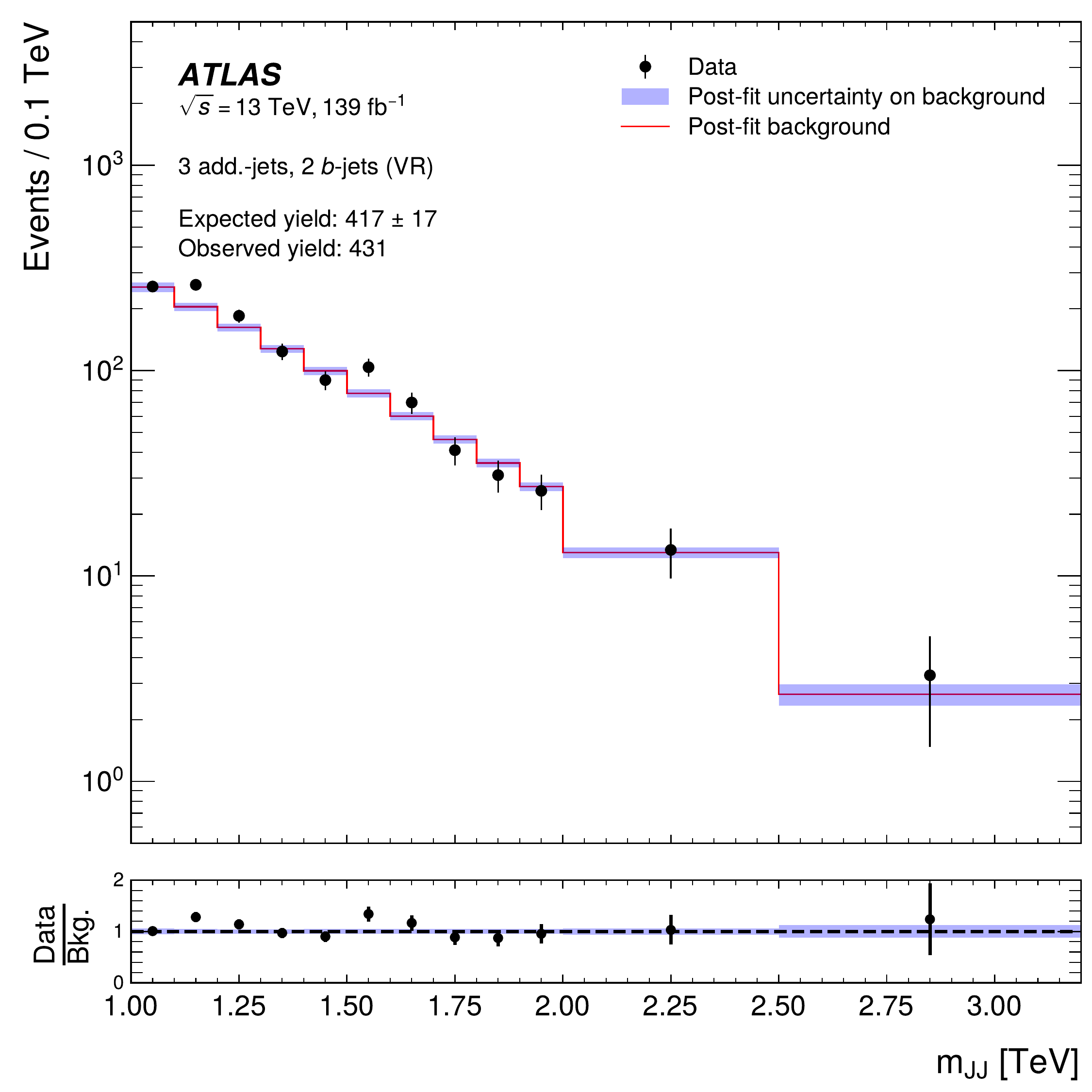}}
\subfloat[]{\includegraphics[width=0.495\textwidth]{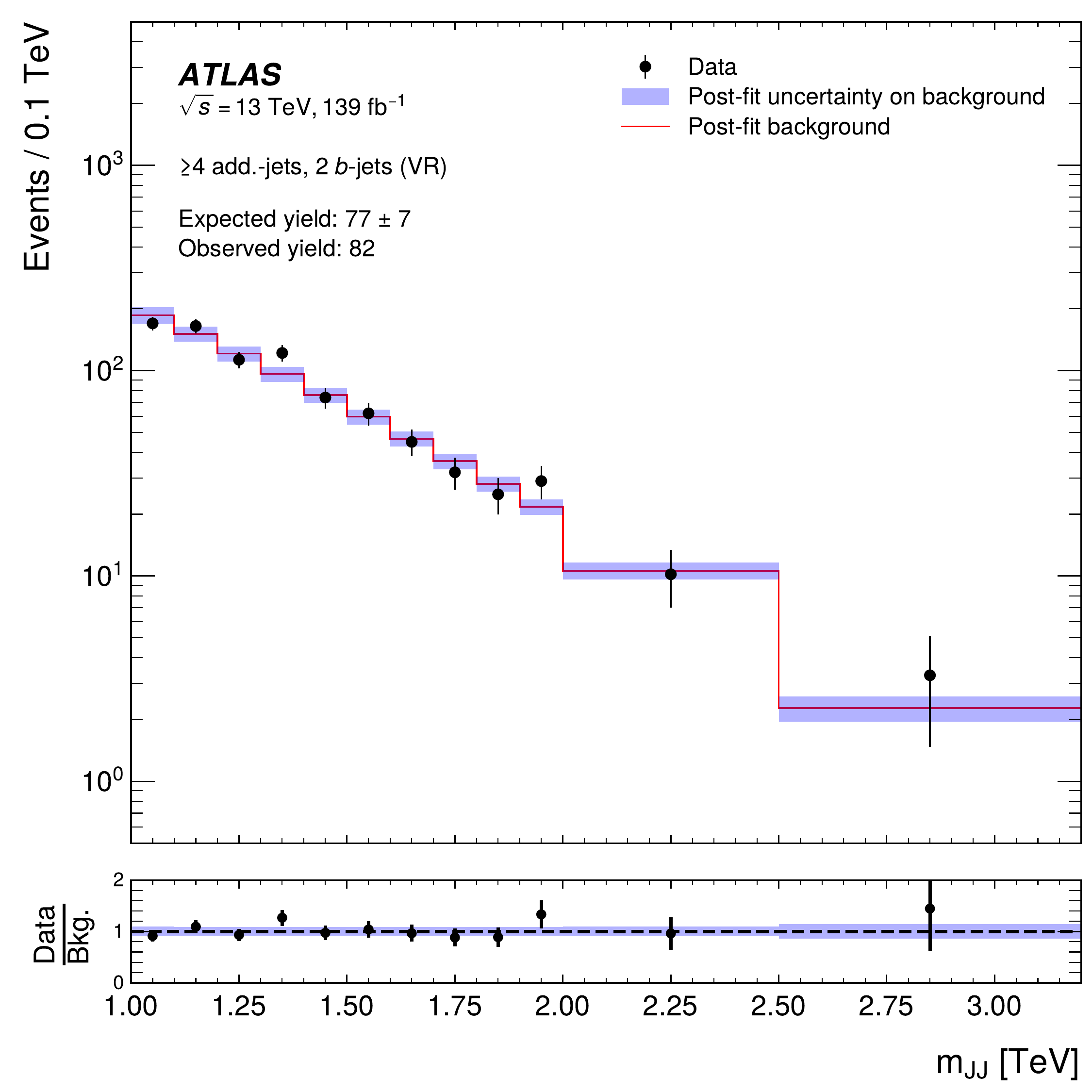}}
\caption{
Comparison of the \mJJ\ distribution between data and the post-fit prediction
under the background-only hypothesis in the validation regions.
Note that the validation regions do not enter the fit.
The post-fit prediction is obtained by propagating the post-fit nuisance parameters from the fit to the validation regions.
The shaded area surrounding the background expectation represents
the total uncertainty from the post-fit computation.
The ratio of the data to the background prediction is shown in the bottom panel.
}
\label{fig:postfit_VR}
\end{figure}
 
The search result with minimal model-dependence is derived based on a simultaneous fit to all signal regions with a background-only hypothesis.
The background estimate is examined in validation regions.
This is done by propagating to the validation regions the post-fit nuisance parameters obtained from the fit to the signal regions.
The resulting agreement between data and the estimated background in the validation regions is shown in Figure~\ref{fig:postfit_VR}.
The largest discrepancies are evaluated and found to be insignificant after taking into account the look elsewhere effect.
Figure~\ref{fig:postfit_BH_1} shows the post-fit \mJJ\ distribution with the background-only hypothesis
compared to the observed data in the two most sensitive signal regions,
along with the results from \textsc{BumpHunter}.
The rest of the signal regions are shown in Figure~\ref{fig:postfit_BH_2}.
All fitted distributions show reasonable agreement with data within uncertainties.
The goodness of fit was evaluated using a likelihood-ratio test
in which the likelihood of the nominal fit is compared to that of a saturated model~\cite{SaturatedModel}.
The result indicates a good description of the data by the background model.
The post-fit nuisance parameter values are all within 1$\sigma$ of the prior uncertainty.
No significant excess was found in the \textsc{BumpHunter} search.
The most significant deviation between data and the background expectation
was observed in the \twoafourb\ region at 1.2~\TeV.
The window with the largest deviation has a local significance of 1.04$\sigma$;
the corresponding global $p$-value is 0.15.
The expected distribution with the presence of a top-philic $Z'$ signal from MC simulation
is illustrated in the top panels of Figure~\ref{fig:postfit_BH_1} and Figure~\ref{fig:postfit_BH_2}.
The top-philic $Z'$ signal considered has $\mZp=1.5$~\TeV\ with  \(c_{t} = 1\) and \(\theta = \pi/2\), and is normalised to have an
arbitrary large cross section of 51~fb, obtained using the cross section predicted by
the simplified model~\cite{PhysRevD.94.035023} with these parameters scaled by a factor of 100.

\begin{figure}[]
\centering
\subfloat[]{\includegraphics[width=0.495\textwidth]{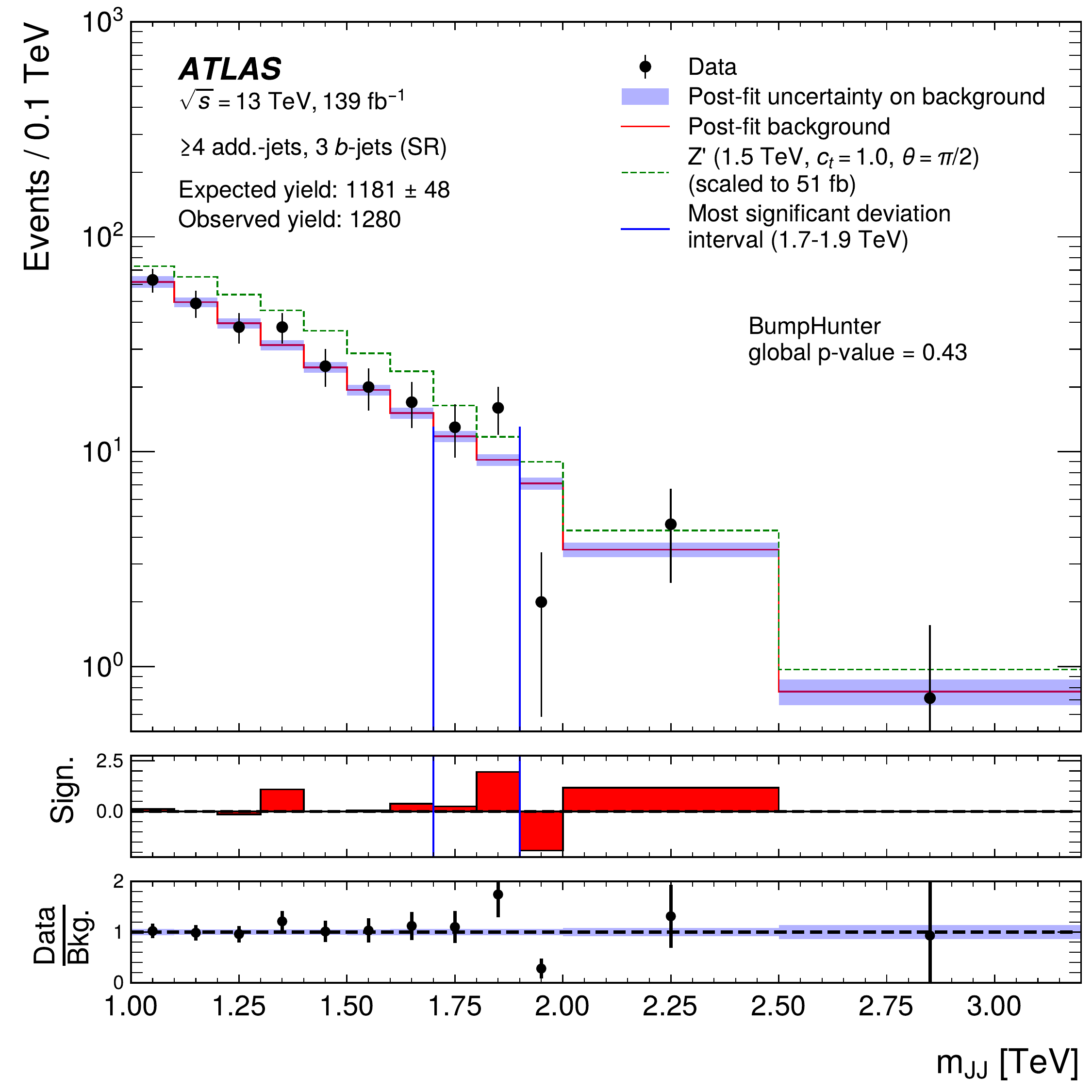}}
\subfloat[]{\includegraphics[width=0.495\textwidth]{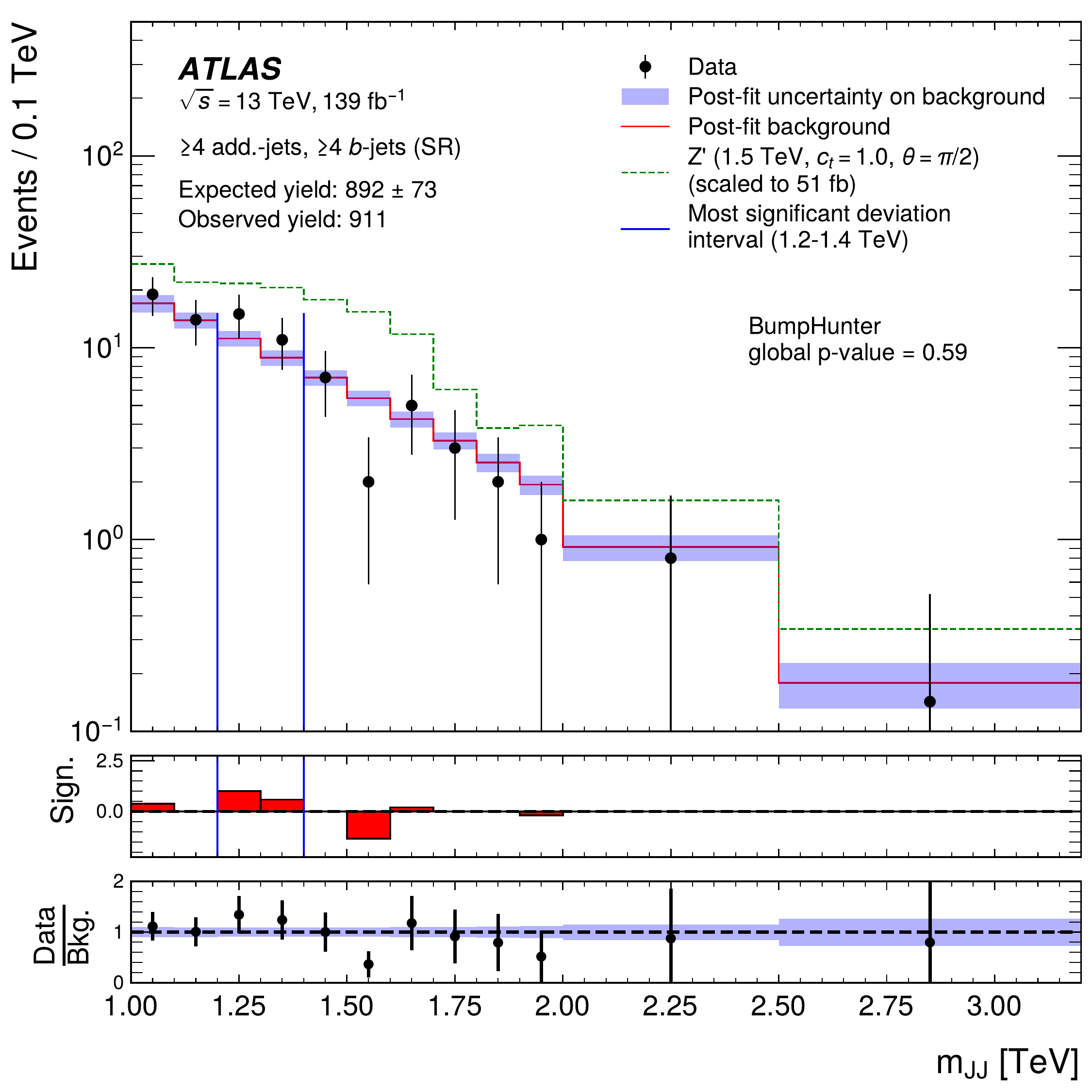}}
\caption{
Comparison of the \mJJ\ distribution between data and the post-fit prediction
under the background-only hypothesis in the two most sensitive signal regions (with four additional jets),
as well as the \textsc{BumpHunter} results.
The shaded area surrounding the background expectation represents
the total uncertainty from the post-fit computation.
The window with the largest deviation between the data and the background expectation
is highlighted in each region.
The local significance of each bin is illustrated in the middle panel.
The ratio of the data to the background prediction is shown in the bottom panel.
The top-philic $Z'$ signal with $\mZp=1.5$~\TeV\ from MC simulation is illustrated in the top panel with dashed lines.
The signal sample is stacked on top of the total background prediction,
and is scaled to have an arbitrary large cross section of 51~fb for visualisation purposes.
}
\label{fig:postfit_BH_1}
\end{figure}

\begin{figure}[]
\centering
\subfloat[]{\includegraphics[width=0.495\textwidth]{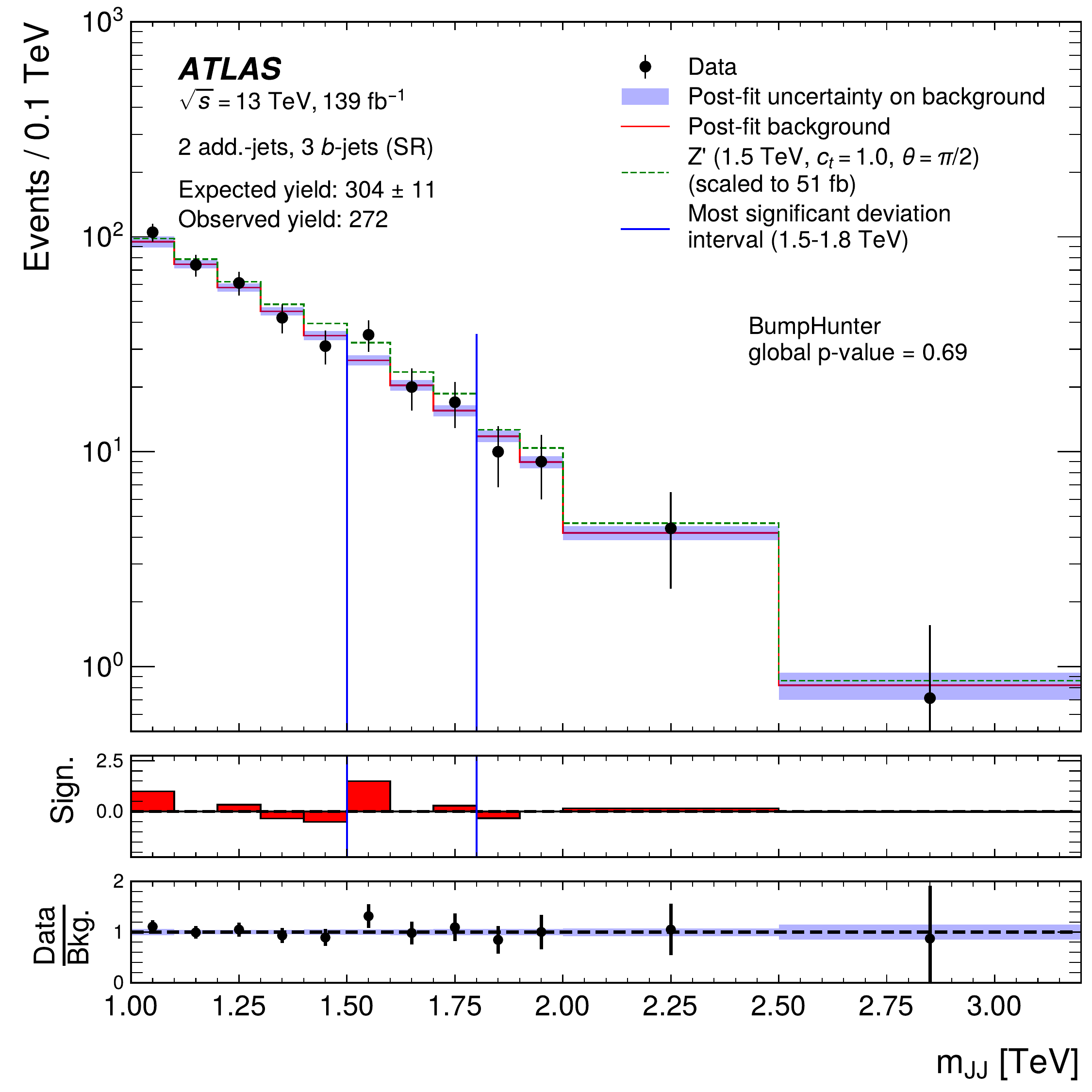}}
\subfloat[]{\includegraphics[width=0.495\textwidth]{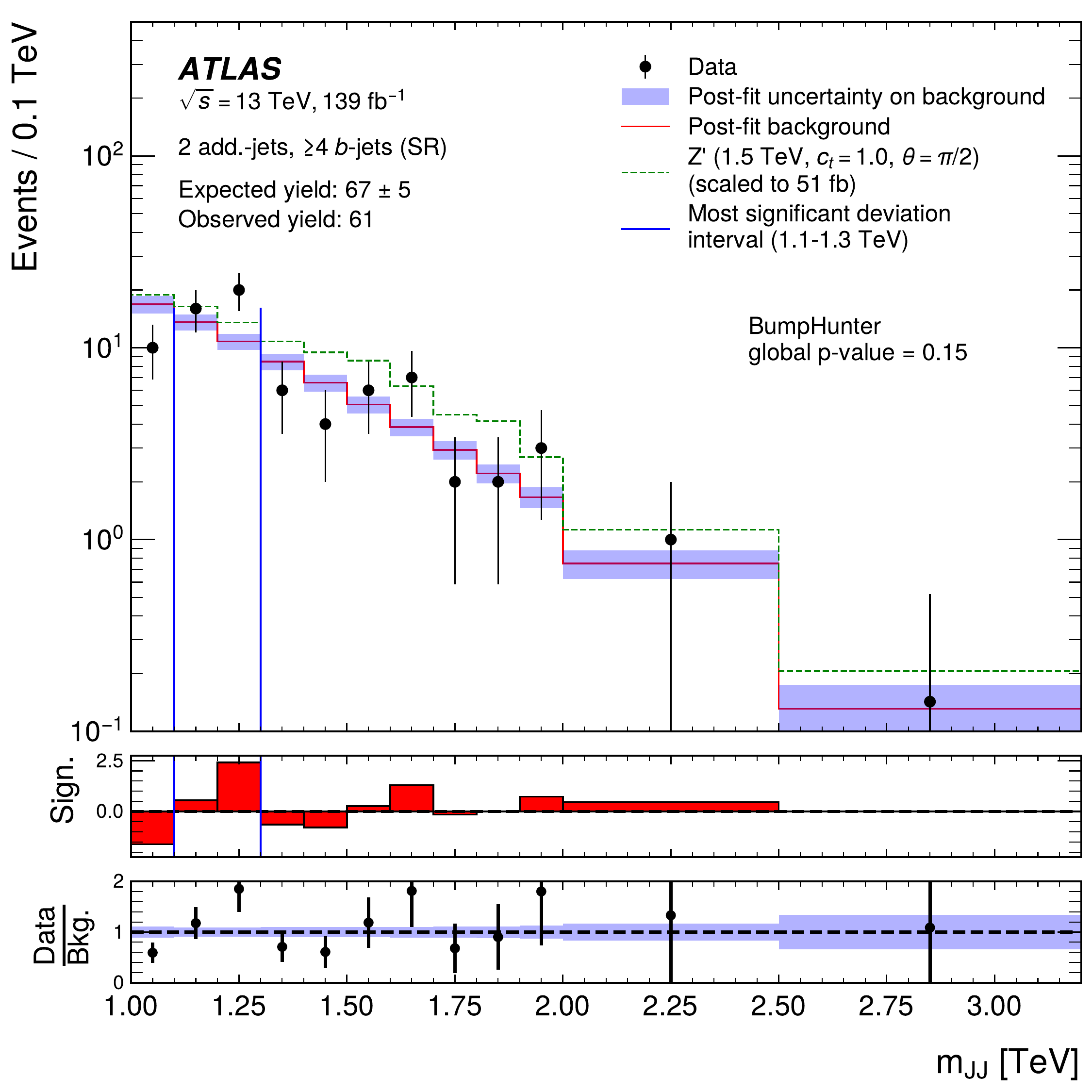}}\\
\subfloat[]{\includegraphics[width=0.495\textwidth]{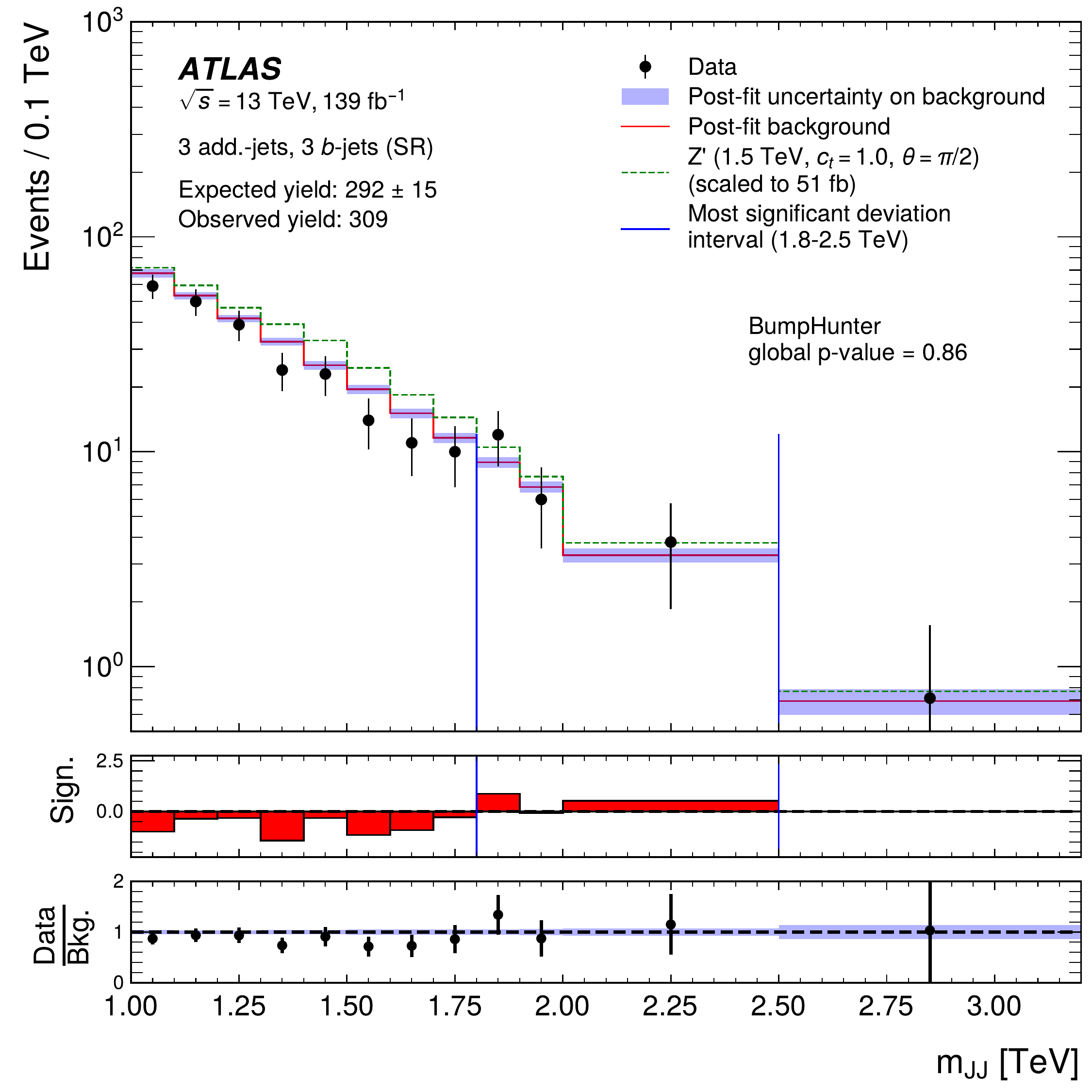}}
\subfloat[]{\includegraphics[width=0.495\textwidth]{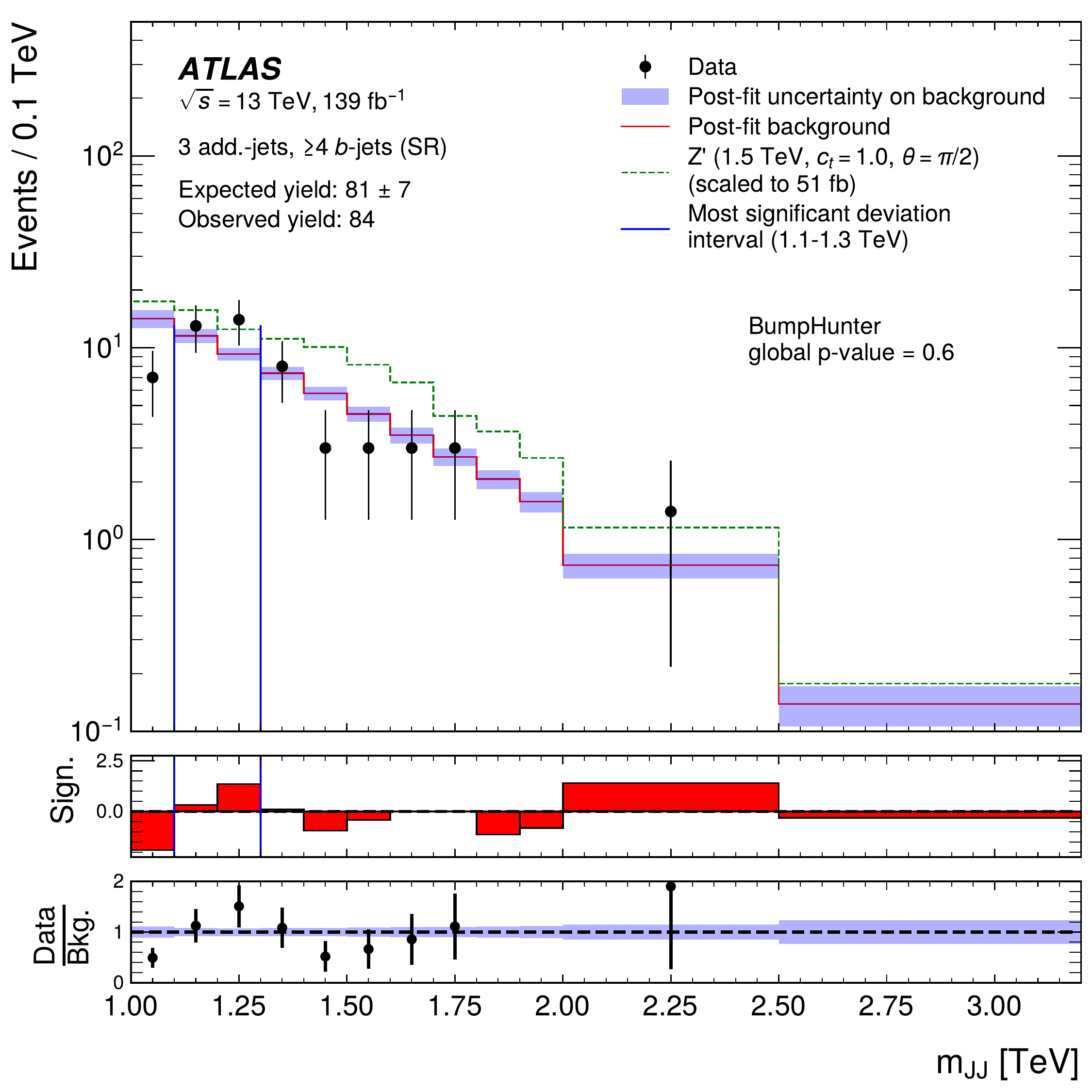}}
\caption{
Comparison of the \mJJ\ distribution between data and the post-fit prediction
under the background-only hypothesis in the signal regions with two or three additional jets.
The shaded area surrounding the background expectation represents
the total uncertainty from the post-fit computation.
The window with the largest deviation between the data and the background expectation
is highlighted in each signal region.
The local significance of each bin is illustrated in the middle panel for the signal regions.
The ratio of the data to the background prediction is shown in the bottom panel.
The top-philic $Z'$ signal with $\mZp=1.5$~\TeV\ from MC simulation is illustrated in the top panel with dashed lines.
The signal sample is stacked on top of the total background prediction,
and is scaled to have an arbitrary large cross section of 51~fb for visualisation purposes.
}
\label{fig:postfit_BH_2}
\end{figure}
 
The model-dependent signal extraction also reveals no significant excess in data.
For all model parameter choices, the fitted signal strength $\mu$ is compatible with zero.
The smallest local $p_0$ value is 3.8\%, obtained with the 1.25~\tev\ mass point for $c_t=1$ and $\theta = 0$.
In addition, the fitted values of the nuisance parameter are compared to those obtained in the fit with the background-only hypothesis
and found to be consistent.
The observed and expected upper limits on the signal production cross section at 95\% confidence level
are shown for specific signal scenarios in Figure~\ref{fig:limits}.
The observed (expected) limits range from 21~(14)~fb to 119~(86)~fb depending on the choice of model parameters. The strongest observed exclusion is observed at \mZp = 2.5 TeV with $c_t = 1$ and $\theta = \pi/2$. At the higher mass range cross section limits are stronger for the smaller couplings $c_t$ due to the narrower signal width. While for the cross section limits there is no strong dependence on the chirality parameter $\theta$, one can observe a stronger model constraint for $\theta = 0$ because of the additional contributions from $tjZ'$ and $tWZ'$ production to the expected signal cross section. 
For $c_t=4$ top-phillic $Z'$ mass of 1.0 TeV is excluded for both chirality parameter choices ($\theta = 0$ and $\pi/2$).
 
\begin{figure}[]
\centering
\subfloat[]{
\includegraphics[width=0.47\textwidth]{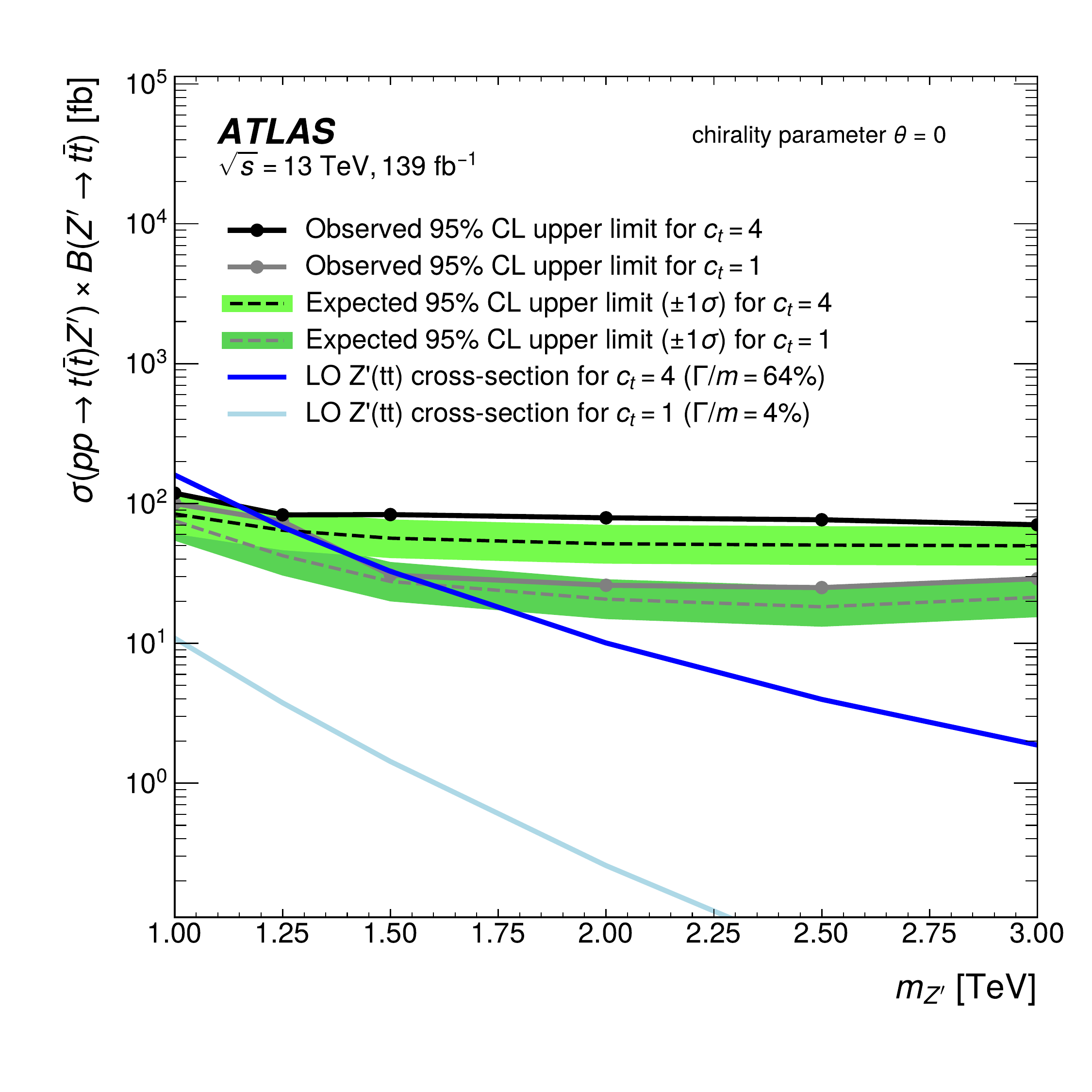}
}
\subfloat[]{
\includegraphics[width=0.47\textwidth]{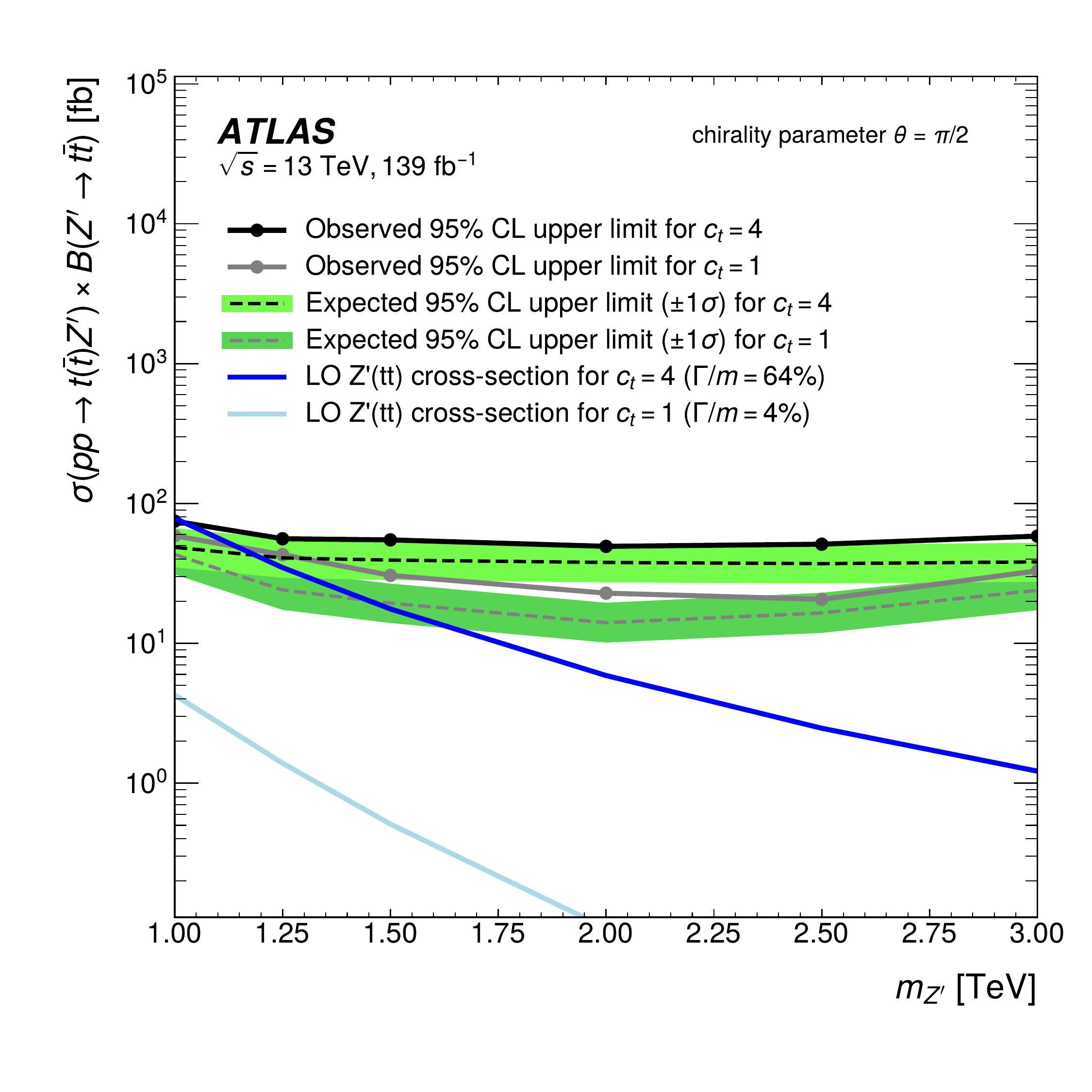}
}
\caption{
Expected and observed 95\% confidence level upper limits on the production cross section of the top-philic $Z'$ signal
with different \mZp\ and for two chirality parameter choices (a) $\theta = 0$ and (b) $\theta =\pi /2$. In each plot two coupling-strength scenarios are shown, for $c_t = 1$ (in gray) and $c_t = 4$ (in black).
The light (dark) blue curves illustrate the signal cross sections based on the model described in Ref.~\cite{PhysRevD.94.035023} for $c_t = 1$ ($c_t = 4$).
The green bands surrounding the expected limits for $c_t = 1$ and $c_t = 4$ correspond to
the 68\% confidence intervals.
All limit lines are obtained by interpolating linearly between the different mass hypotheses.
}
\label{fig:limits}
\end{figure}

Table~\ref{tab:impact_table} lists the impacts of various groups of uncertainties
relative to the total uncertainty on the fitted signal strength
for two examples of the top-philic $Z'$ signal: ($\mZp=1.5~\TeV$, $c_t = 1$, $\theta=0$) and ($\mZp=3.0~\TeV$, $c_t=4$, $\theta=0$).
The results are dominated by systematic uncertainties.
The most important source of systematic uncertainty is the modelling of the \ttjets\ background.
The uncertainties due to the migration and acceptance effect of the ME and PS generator choices in the ${\ge}4b$ regions play a dominant role.
They contribute 26\% and 31\% (36\% and 48\%) of the total uncertainty, respectively,
for the signal with $\mZp=1.5~\TeV$ and $c_t = 1$ ($\mZp=3.0~\TeV$ and $c_t=4$).
This reflects the large difference between the predictions of the different generators in the phase space explored by this search.
The uncertainties in the normalisation of \ttbjets\ also contribute 18\% of the total uncertainty for both signal hypotheses shown in Table~~\ref{tab:impact_table}.
The uncertainties from the jet energy scale and resolution are also an important source,
given the large jet multiplicity in the signal regions.
This is followed by the uncertainties from the data-driven background estimate.

\begin{table}[]
\centering
\caption{
The contribution from different systematic uncertainties
relative to the total uncertainty on the fitted signal strength for two signal scenarios with $\theta =0$, grouped into categories.
For each category, the fit is repeated with the corresponding group of nuisance parameters
fixed to their best-fit values.
The contribution from each category is then evaluated
by subtracting in quadrature the uncertainty on the signal strength obtained in this fit
from that of the full fit with all uncertainties.
The percentage is calculated relative to the total uncertainty from the full fit.
The contribution from the statistical uncertainty is also shown.
The total systematic uncertainty is different from the sum in quadrature of the different groups
due to the correlations among the nuisance parameters in the fit.
\newline
}
\setlength\extrarowheight{3pt}
\begin{tabular}{l >{\centering}m{0.125\textwidth} >{\centering\arraybackslash}m{0.125\textwidth} }
\toprule
\midrule
\multirow{2}{*}{Uncertainty categories} & \multicolumn{2}{>{\centering}m{0.25\textwidth}}{Relative contribution to the total uncertainty [\%]} \\ \cline{2-3}
& 1.5~\TeV \; $c_t=1$ & 3~\TeV \; $c_t = 4$ \\
\midrule
\ttjets\ modelling                         & 78       & 87     \\
Jet energy scale and resolution           & 40       & 46     \\
Functional fit and extrapolation          & 41       & 23     \\
Signal bias                               & 35      & 7.3    \\
Statistical uncertainty on signal MC      & 26      & 19     \\
Single-top-quark modelling                 & 11      & 8.9     \\
Flavour tagging                            & 9.4      & 9.8    \\
Minor backgrounds modelling                & 6.9      & 5.3    \\
Other uncertainties                       & 1.5      & 2.1    \\
Luminosity                                & 0.2      & 2.0    \\
\midrule
Total systematic uncertainty              & 95       & 96     \\
Statistical uncertainty                   & 32       & 29     \\
\midrule
\bottomrule
\end{tabular}
\label{tab:impact_table}
\end{table}
 
\FloatBarrier

% End of text imported from the .//results/results.tex input file

\FloatBarrier

% The next lines are included from the .//conclusion/conclusion.tex input file
\section{Conclusion}
\label{sec:conclusion}
 
A search for a top-philic heavy resonance produced in association with a top quark or a top-quark pair and decaying to \ttbar\ is presented.
The search is performed using 139~\ifb\ of $pp$ collision data at $\sqrt{s} = 13~\TeV$
collected by the ATLAS detector at the LHC.
Large-$R$ jets are used as proxies of the top quarks from the heavy resonance decay.
The invariant mass spectra of the two large-$R$ jets with the largest \pt\
are used to test for the presence of a resonance signal in the range of 1.0~\TeV\ to 3.2~\TeV.
Events in the single-lepton final state are selected
and categorised according to the number of additional jets and $b$-tagged jets.
The search is conducted in regions with at least three $b$-tagged jets,
with and without assuming a specific $Z'$ signal model.
No significant excess was observed above the expected background.
The upper limits on the $Z'$ production cross section at 95\% confidence level are computed
for signals with six values of \mZp\ between 1.0~\TeV\ and 3.0~\TeV\ based on a simplified model.
The observed (expected) limits range from 21~(14)~fb to 119~(86)~fb depending on the choice of model parameters.

% End of text imported from the .//conclusion/conclusion.tex input file

\section*{Acknowledgements} 

% The next lines are included from the .//acknowledgements/Acknowledgements.tex input file

We thank CERN for the very successful operation of the LHC, as well as the
support staff from our institutions without whom ATLAS could not be
operated efficiently.
 
We acknowledge the support of
ANPCyT, Argentina;
YerPhI, Armenia;
ARC, Australia;
BMWFW and FWF, Austria;
ANAS, Azerbaijan;
CNPq and FAPESP, Brazil;
NSERC, NRC and CFI, Canada;
CERN;
ANID, Chile;
CAS, MOST and NSFC, China;
Minciencias, Colombia;
MEYS CR, Czech Republic;
DNRF and DNSRC, Denmark;
IN2P3-CNRS and CEA-DRF/IRFU, France;
SRNSFG, Georgia;
BMBF, HGF and MPG, Germany;
GSRI, Greece;
RGC and Hong Kong SAR, China;
ISF and Benoziyo Center, Israel;
INFN, Italy;
MEXT and JSPS, Japan;
CNRST, Morocco;
NWO, Netherlands;
RCN, Norway;
MEiN, Poland;
FCT, Portugal;
MNE/IFA, Romania;
MESTD, Serbia;
MSSR, Slovakia;
ARRS and MIZ\v{S}, Slovenia;
DSI/NRF, South Africa;
MICINN, Spain;
SRC and Wallenberg Foundation, Sweden;
SERI, SNSF and Cantons of Bern and Geneva, Switzerland;
MOST, Taiwan;
TENMAK, T\"urkiye;
STFC, United Kingdom;
DOE and NSF, United States of America.
In addition, individual groups and members have received support from
BCKDF, CANARIE, Compute Canada and CRC, Canada;
PRIMUS 21/SCI/017 and UNCE SCI/013, Czech Republic;
COST, ERC, ERDF, Horizon 2020 and Marie Sk{\l}odowska-Curie Actions, European Union;
Investissements d'Avenir Labex, Investissements d'Avenir Idex and ANR, France;
DFG and AvH Foundation, Germany;
Herakleitos, Thales and Aristeia programmes co-financed by EU-ESF and the Greek NSRF, Greece;
BSF-NSF and MINERVA, Israel;
Norwegian Financial Mechanism 2014-2021, Norway;
NCN and NAWA, Poland;
La Caixa Banking Foundation, CERCA Programme Generalitat de Catalunya and PROMETEO and GenT Programmes Generalitat Valenciana, Spain;
G\"{o}ran Gustafssons Stiftelse, Sweden;
The Royal Society and Leverhulme Trust, United Kingdom.
 
The crucial computing support from all WLCG partners is acknowledged gratefully, in particular from CERN, the ATLAS Tier-1 facilities at TRIUMF (Canada), NDGF (Denmark, Norway, Sweden), CC-IN2P3 (France), KIT/GridKA (Germany), INFN-CNAF (Italy), NL-T1 (Netherlands), PIC (Spain), ASGC (Taiwan), RAL (UK) and BNL (USA), the Tier-2 facilities worldwide and large non-WLCG resource providers. Major contributors of computing resources are listed in Ref.~\cite{ATL-SOFT-PUB-2021-003}.

% End of text imported from the .//acknowledgements/Acknowledgements.tex input file

\clearpage
 
\printbibliography

@article{Greiner:2014qna,
    author = "Greiner, Nicolas and Kong, Kyoungchul and Park, Jong-Chul and Park, Seong Chan and Winter, Jan-Christopher",
    title = "{Model-Independent Production of a Top-Philic Resonance at the LHC}",
    eprint = "1410.6099",
    archivePrefix = "arXiv",
    primaryClass = "hep-ph",
    reportNumber = "DESY-14-177, MPP-2014-371",
    doi = "10.1007/JHEP04(2015)029",
    journal = "JHEP",
    volume = "04",
    pages = "029",
    year = "2015"
}

@inproceedings{Choudalakis:2011qn,
      author         = "Choudalakis, Georgios",
      title          = "{On hypothesis testing, trials factor, hypertests and the
                        BumpHunter}",
      booktitle      = "{Proceedings, PHYSTAT 2011 Workshop on Statistical Issues
                        Related to Discovery Claims in Search Experiments and
                        Unfolding, CERN,Geneva, Switzerland 17-20 January 2011}",
      year           = "2011",
      eprint         = "1101.0390",
      archivePrefix  = "arXiv",
      primaryClass   = "physics.data-an",
      SLACcitation   = "%%CITATION = ARXIV:1101.0390;%%"
}

@article{LUCID2,
  author={G. Avoni and others},
  title={The new LUCID-2 detector for luminosity measurement and monitoring in ATLAS},
  journal={JINST},
  volume={13},
  number={07},
  pages={P07017},
  doi="10.1088/1748-0221/13/07/P07017",
  year={2018},
}

@misc{SaturatedModel,
author = "{Robert D. Cousins}",
title = "{Generalization of Chisquare Goodness-of-Fit Test for Binned Data Using Saturated Models, with Application to Histograms}",
url={http://www.physics.ucla.edu/~cousins/stats/cousins_saturated.pdf},
year = 2013
}

@article{Ferretti:2013kya,
      author         = "Ferretti, Gabriele and Karateev, Denis",
      title          = "{Fermionic UV completions of Composite Higgs models}",
      journal        = "JHEP",
      volume         = "03",
      year           = "2014",
      pages          = "077",
      doi            = "10.1007/JHEP03(2014)077",
      eprint         = "1312.5330",
      archivePrefix  = "arXiv",
      primaryClass   = "hep-ph",
      SLACcitation   = "%%CITATION = ARXIV:1312.5330;%%"
}

@article{Vecchi:2015fma,
      author         = "Vecchi, Luca",
      title          = "{A dangerous irrelevant UV-completion of the composite
                        Higgs}",
      journal        = "JHEP",
      volume         = "02",
      year           = "2017",
      pages          = "094",
      doi            = "10.1007/JHEP02(2017)094",
      eprint         = "1506.00623",
      archivePrefix  = "arXiv",
      primaryClass   = "hep-ph",
      SLACcitation   = "%%CITATION = ARXIV:1506.00623;%%"
}

@article{Agashe:2003zs,
      author         = "Agashe, Kaustubh and Delgado, Antonio and May, Michael J.
                        and Sundrum, Raman",
      title          = "{RS1, custodial isospin and precision tests}",
      journal        = "JHEP",
      volume         = "08",
      year           = "2003",
      pages          = "050",
      doi            = "10.1088/1126-6708/2003/08/050",
      eprint         = "hep-ph/0308036",
      archivePrefix  = "arXiv",
      primaryClass   = "hep-ph",
      SLACcitation   = "%%CITATION = HEP-PH/0308036;%%"
}

@article{Agashe:2004rs,
      author         = "Agashe, Kaustubh and Contino, Roberto and Pomarol, Alex",
      title          = "{The Minimal composite Higgs model}",
      journal        = "Nucl. Phys.",
      volume         = "B719",
      year           = "2005",
      pages          = "165-187",
      doi            = "10.1016/j.nuclphysb.2005.04.035",
      eprint         = "hep-ph/0412089",
      archivePrefix  = "arXiv",
      primaryClass   = "hep-ph",
      reportNumber   = "UAB-FT-567",
      SLACcitation   = "%%CITATION = HEP-PH/0412089;%%"
}

@article{PhysRevD.94.035023,
    author = "Kim, Jeong Han and Kong, Kyoungchul and Lee, Seung J. and Mohlabeng, Gopolang",
    title = "{Probing TeV scale Top-Philic Resonances with Boosted Top-Tagging at the High Luminosity LHC}",
    eprint = "1604.07421",
    archivePrefix = "arXiv",
    primaryClass = "hep-ph",
    reportNumber = "FERMILAB-PUB-16-125-T",
    doi = "10.1103/PhysRevD.94.035023",
    journal = "Phys. Rev. D",
    volume = "94",
    number = "3",
    pages = "035023",
    year = "2016"
}

@article{Aad:2020gmm,
    author = "ATLAS Collaboration",
    collaboration = "ATLAS",
    title = "{Muon reconstruction and identification efficiency in ATLAS using the full Run 2 $pp$ collision data set at $\sqrt{s}=13$ TeV}",
    eprint = "2012.00578",
    archivePrefix = "arXiv",
    primaryClass = "hep-ex",
    reportNumber = "CERN-EP-2020-199",
    doi = "10.1140/epjc/s10052-021-09233-2",
    journal = "Eur. Phys. J. C",
    volume = "81",
    number = "7",
    pages = "578",
    year = "2021"
}

@article{JETM-2018-05,
    author = "ATLAS Collaboration",
    collaboration = "ATLAS",
    title = "{Jet energy scale and resolution measured in proton\textendash{}proton collisions at $\sqrt{s}=13$~TeV with the ATLAS detector}",
    eprint = "2007.02645",
    archivePrefix = "arXiv",
    primaryClass = "hep-ex",
    reportNumber = "CERN-EP-2020-083",
    doi = "10.1140/epjc/s10052-021-09402-3",
    journal = "Eur. Phys. J. C",
    volume = "81",
    number = "8",
    pages = "689",
    year = "2021"
}

@article{Krohn:2009th,
    author = "Krohn, David and Thaler, Jesse and Wang, Lian-Tao",
    title = "{Jet Trimming}",
    eprint = "0912.1342",
    archivePrefix = "arXiv",
    primaryClass = "hep-ph",
    doi = "10.1007/JHEP02(2010)084",
    journal = "JHEP",
    volume = "02",
    pages = "084",
    year = "2010"
}

@article{Gross_2010,
    author = "Gross, Eilam and Vitells, Ofer",
    title = "{Trial factors for the look elsewhere effect in high energy physics}",
    eprint = "1005.1891",
    archivePrefix = "arXiv",
    primaryClass = "physics.data-an",
    doi = "10.1140/epjc/s10052-010-1470-8",
    journal = "Eur. Phys. J. C",
    volume = "70",
    pages = "525--530",
    year = "2010"
}

@techreport{Cranmer:1456844,
      author        = "Cranmer, Kyle and Lewis, George and Moneta, Lorenzo and
                       Shibata, Akira and Verkerke, Wouter",
      title         = "{HistFactory: A tool for creating statistical models for
                       use with RooFit and RooStats}",
      institution   = "New York U.",
      collaboration = "ROOT Collaboration",
      address       = "New York",
      reportNumber  = "CERN-OPEN-2012-016",
      month         = "Jan",
      year          = "2012",
      url           = "https://cds.cern.ch/record/1456844",
}

@misc{moneta2011roostats,
      title={The RooStats Project}, 
      author={Lorenzo Moneta and Kevin Belasco and Kyle Cranmer and Sven Kreiss and Alfio Lazzaro and Danilo Piparo and Gregory Schott and Wouter Verkerke and Matthias Wolf},
      year={2011},
      eprint={1009.1003},
      archivePrefix={arXiv},
      primaryClass={physics.data-an}
}

@article{Junk_1999,
    author = "Junk, Thomas",
    title = "{Confidence level computation for combining searches with small statistics}",
    eprint = "hep-ex/9902006",
    archivePrefix = "arXiv",
    reportNumber = "CARLETON-OPAL-PHYS-99-01, CERN-EP-99-041",
    doi = "10.1016/S0168-9002(99)00498-2",
    journal = "Nucl. Instrum. Meth. A",
    volume = "434",
    pages = "435--443",
    year = "1999"
}

@article{Aad:2020klt,
    author         = "{ATLAS Collaboration}",
    title = "{Evidence for $t\bar{t}t\bar{t}$ production in the multilepton final state in proton\textendash{}proton collisions at $\sqrt{s}=13$ $\text {TeV}$ with the ATLAS detector}",
    eprint = "2007.14858",
    archivePrefix = "arXiv",
    primaryClass = "hep-ex",
    reportNumber = "CERN-EP-2020-111",
    doi = "10.1140/epjc/s10052-020-08509-3",
    journal = "Eur. Phys. J. C",
    volume = "80",
    number = "11",
    pages = "1085",
    year = "2020"
}

@article{Alwall:2008epjc,
    author         = "Alwall, J. and Hoeche, S. and Krauss, F. and
    Lavesson, N. and Loennblad, L. and Maltoni, F. and Mangano, M.A. and Moretti, M. and
    Papadopoulos, C.G. and Piccinini, F. and Schumann, S. and Treccani, M. and Winter, J.
    and Worek, M.",
    title = "{Comparative study of various algorithms for the merging of parton showers
    and matrix elements in hadronic collisions}",
    eprint = "0706.2569",
    archivePrefix = "arXiv",
    primaryClass = "hep-ph",
    doi = "10.1140/epjc/s10052-007-0490-5",
    journal = "Eur. Phys. J. C",
    volume = "53",
    pages = "473",
    year = "2008"
}

@article{ATLAS:2021kqb,
    author = "ATLAS Collaboration",
    collaboration = "ATLAS",
    title = "{Measurement of the t$ \overline{t} $t$ \overline{t} $ production cross section in $pp$ collisions at $ \sqrt{s} $ = 13 TeV with the ATLAS detector}",
    eprint = "2106.11683",
    archivePrefix = "arXiv",
    primaryClass = "hep-ex",
    reportNumber = "CERN-EP-2021-075",
    doi = "10.1007/JHEP11(2021)118",
    journal = "JHEP",
    volume = "11",
    pages = "118",
  year = "2021"
}

@misc{Frixione:2007nu,
    author = "Frixione, Stefano and Nason, Paolo and Ridolfi, Giovanni",
    title = "{The POWHEG-hvq manual version 1.0}",
    eprint = "0707.3081",
    archivePrefix = "arXiv",
    primaryClass = "hep-ph",
    reportNumber = "BICOCCA-FT-07-10, GEF-TH-16-2007",
    month = "7",
    year = "2007"
}

@article{Sirunyan:2019nxl,
      author         = "{CMS Collaboration}",
      title          = "{Search for the production of four top quarks in the single-lepton and opposite-sign dilepton final states in proton-proton collisions at \(\sqrt{s} = 13\,\text{TeV}\)}",
      journal        = "JHEP",
      volume         = "11",
      year           = "2019",
      pages          = "082",
      doi            = "10.1007/JHEP11(2019)082",
      reportNumber   = "CERN-EP-2019-098",
      eprint         = "1906.02805",
      archivePrefix  = "arXiv",
      primaryClass   = "hep-ex",
      SLACcitation   = "%%CITATION = ARXIV:1906.02805;%%"
}

@article{Mattelaer:2016gcx,
    author = "Mattelaer, Olivier",
    title = "{On the maximal use of Monte Carlo samples: re-weighting events at NLO accuracy}",
    eprint = "1607.00763",
    archivePrefix = "arXiv",
    primaryClass = "hep-ph",
    reportNumber = "IPPP-16-55, MCNET-16-25",
    doi = "10.1140/epjc/s10052-016-4533-7",
    journal = "Eur. Phys. J. C",
    volume = "76",
    number = "12",
    pages = "674",
    year = "2016"
}

@Report{ATLAS-TDR-19,
    author         = "{ATLAS Collaboration}",
    title          = "{ATLAS Insertable B-Layer: Technical Design Report}",
    type           = "ATLAS-TDR-19; CERN-LHCC-2010-013",
    year           = "2010",
    url            = "https://cds.cern.ch/record/1291633",
    related        = "ATLAS-TDR-19-addm",
    relatedstring  = "Addendum:",
}

@Article{PIX-2018-001,
    author         = "Abbott, B. and others",
    title          = "{Production and integration of the ATLAS Insertable B-Layer}",
    journal        = "JINST",
    volume         = "13",
    year           = "2018",
    pages          = "T05008",
    doi            = "10.1088/1748-0221/13/05/T05008",
    eprint         = "1803.00844",
    archivePrefix  = "arXiv",
    primaryClass   = "physics.ins-det",
}

@article{ATLAS:2022qxm,
    author         = "{ATLAS Collaboration}",
    collaboration = "ATLAS",
    title = "{ATLAS flavour-tagging algorithms for the LHC Run 2 $pp$ collision dataset}",
    eprint = "2211.16345",
    archivePrefix = "arXiv",
    primaryClass = "physics.data-an",
    reportNumber = "CERN-EP-2022-226",
    month = "11",
    year = "2022"
}

@article{Bothmann2019,
  doi = {10.21468/scipostphys.7.3.034},
  url = {https://doi.org/10.21468/scipostphys.7.3.034},
  year = {2019},
  month = sep,
  publisher = {Stichting {SciPost}},
  volume = {7},
  number = {3},
  author = {Enrico Bothmann and Gurpreet Singh Chahal and Stefan H\"{o}che and Johannes Krause and Frank Krauss and Silvan Kuttimalai and Sebastian Liebschner and Davide Napoletano and Marek Sch\"{o}nherr and Holger Schulz and Steffen Schumann and Frank Siegert},
  title = {Event generation with Sherpa 2.2},
  journal = {{SciPost} Physics}
}

@article{Martin2009,
  doi = {10.1140/epjc/s10052-009-1072-5},
  url = {https://doi.org/10.1140/epjc/s10052-009-1072-5},
  year = {2009},
  month = jul,
  publisher = {Springer Science and Business Media {LLC}},
  volume = {63},
  number = {2},
  pages = {189--285},
  author = {A. D. Martin and W. J. Stirling and R. S. Thorne and G. Watt},
  title = {Parton distributions for the {LHC}},
  journal = {The European Physical Journal C}
}

@Article{FTAG-2019-02,
    author         = "{ATLAS Collaboration}",
    title          = "{Calibration of the light-flavour jet mistagging efficiency of the \(b\)-tagging algorithms with \(Z\)+jets events using \(139\,\text{fb}^{-1}\) of ATLAS proton--proton collision data at \(\sqrt{s} = 13\,\text{TeV}\)}",
    journal        = "Eur. Phys. J. C",
    volume         = "83",
    year           = "2023",
    pages          = "728",
    doi            = "10.1140/epjc/s10052-023-11736-z",
    reportNumber   = "CERN-EP-2022-211",
    eprint         = "2301.06319",
    archivePrefix  = "arXiv",
    primaryClass   = "hep-ex",
}

@Article{CMS-TOP-21-005,
    author         = "{CMS Collaboration}",
    title          = "{Evidence for four-top quark production in proton--proton collisions at \(\sqrt{s} = 13\,\text{TeV}\)}",
    journal        = "Phys. Lett. B",
    volume         = "844",
    year           = "2023",
    pages          = "138076",
    doi            = "10.1016/j.physletb.2023.138076",
    reportNumber   = "CERN-EP-2023-014",
    eprint         = "2303.03864",
    archivePrefix  = "arXiv",
    primaryClass   = "hep-ex",
}

@Article{TOPQ-2021-08,
    author         = "{ATLAS Collaboration}",
    title          = "{Observation of four-top-quark production in the multilepton final state with the ATLAS detector}",
    journal        = "Eur. Phys. J. C",
    volume         = "83",
    year           = "2023",
    pages          = "496",
    doi            = "10.1140/epjc/s10052-023-11573-0",
    reportNumber   = "CERN-EP-2023-055",
    eprint         = "2303.15061",
    archivePrefix  = "arXiv",
    primaryClass   = "hep-ex",
}

@Article{CMS-TOP-22-013,
    author         = "{CMS Collaboration}",
    title          = "{Observation of four top quark production in proton--proton collisions at \(\sqrt{s} = 13\,\text{TeV}\)}",
    journal        = "Phys. Lett. B",
    volume         = "847",
    year           = "2023",
    pages          = "138290",
    doi            = "10.1016/j.physletb.2023.138290",
    reportNumber   = "CERN-EP-2023-090",
    eprint         = "2305.13439",
    archivePrefix  = "arXiv",
    primaryClass   = "hep-ex",
}

@article{Barger:2010uw,
    author = "Barger, Vernon and Keung, Wai-Yee and Yencho, Brian",
    title = "{Triple-top signal of new physics at the LHC}",
    eprint = "1001.0221",
    archivePrefix = "arXiv",
    primaryClass = "hep-ph",
    doi = "10.1016/j.physletb.2010.03.001",
    journal = "Phys. Lett. B",
    volume = "687",
    pages = "70",
    year = "2010"
}

@article{Chen:2014,
    author = "Chen, Chuan-Ren",
    title = "{Searching for new physics with triple-top signal at the LHC}",
    doi = "10.1016/j.physletb.2014.07.041",
    journal = "Phys. Lett. B",
    volume = "736",
    pages = "321",
    year = "2014"
}

@article{Boos:2021yat,
    author = "Boos, Eduard and Dudko, Lev",
    title = "{Triple top quark production in standard model}",
    eprint = "2107.07629",
    archivePrefix = "arXiv",
    primaryClass = "hep-ph",
    doi = "10.1142/S0217751X22500233",
    journal = "Int. J. Mod. Phys. A",
    volume = "37",
    number = "05",
    pages = "2250023",
    year = "2022"
}

@Article{Cacciari:2008gp,
     author    = "Cacciari, Matteo and Salam, Gavin P. and Soyez, Gregory",
     title     = "{The anti-\(k_{t}\) jet clustering algorithm}",
     journal   = "JHEP",
     volume    = "04",
     year      = "2008",
     pages     = "063",
     eprint    = "0802.1189",
     archivePrefix = "arXiv",
     primaryClass  =  "hep-ph",
     doi       = "10.1088/1126-6708/2008/04/063",
     SLACcitation  = "%%CITATION = 0802.1189;%%"
}

@Article{Fastjet,
      author         = "Cacciari, Matteo and Salam, Gavin P. and Soyez, Gregory",
      title          = "{FastJet user manual}",
      journal        = "Eur. Phys. J. C",
      volume         = "72",
      year           = "2012",
      pages          = "1896",
      doi            = "10.1140/epjc/s10052-012-1896-2",
      eprint         = "1111.6097",
      archivePrefix  = "arXiv",
      primaryClass   = "hep-ph",
      reportNumber   = "CERN-PH-TH-2011-297",
      SLACcitation   = "%%CITATION = ARXIV:1111.6097;%%"
}

@Article{Butterworth:2015oua,
      author         = "Butterworth, Jon and others",
      title          = "{PDF4LHC recommendations for LHC Run II}",
      journal        = "J. Phys. G",
      volume         = "43",
      year           = "2016",
      pages          = "023001",
      doi            = "10.1088/0954-3899/43/2/023001",
      eprint         = "1510.03865",
      archivePrefix  = "arXiv",
      primaryClass   = "hep-ph",
      reportNumber   = "OUTP-15-17P, SMU-HEP-15-12, TIF-UNIMI-2015-14,
                        LCTS-2015-27, CERN-PH-TH-2015-249",
      SLACcitation   = "%%CITATION = ARXIV:1510.03865;%%"
}

@article{Harland-Lang:2014zoa,
      author         = "Harland-Lang, L. A. and Martin, A. D. and Motylinski, P.
                        and Thorne, R. S.",
      title          = "{Parton distributions in the LHC era: MMHT 2014 PDFs}",
      journal        = "Eur. Phys. J. C",
      number         = "5",
      volume         = "75",
      pages          = "204",
      doi            = "10.1140/epjc/s10052-015-3397-6",
      year           = "2015",
      eprint         = "1412.3989",
      archivePrefix  = "arXiv",
      primaryClass   = "hep-ph",
      reportNumber   = "LCTS-2014-47, IPPP-14-97, DCPT-14-194",
      SLACcitation   = "%%CITATION = ARXIV:1412.3989;%%",
}

@article{Ball:2014uwa,
      author         = "Ball, Richard D. and others",
      title          = "{Parton distributions for the LHC run II}",
      collaboration  = "NNPDF",
      journal        = "JHEP",
      volume         = "04",
      year           = "2015",
      pages          = "040",
      doi            = "10.1007/JHEP04(2015)040",
      eprint         = "1410.8849",
      archivePrefix  = "arXiv",
      primaryClass   = "hep-ph",
      reportNumber   = "EDINBURGH-2014-15, IFUM-1034-FT, CERN-PH-TH-2013-253,
                        OUTP-14-11P, CAVENDISH-HEP-14-11",
      SLACcitation   = "%%CITATION = ARXIV:1410.8849;%%"
}

@Article{Sjostrand:2007gs,
	author    = "Sj{\"o}strand, T. and Mrenna, S. and Skands, P.",
	title     = "{A brief introduction to PYTHIA 8.1}",
	journal   = "Comput. Phys. Commun.",
	volume    = "178",
	year      = "2008",
	pages     = "852-867",
	eprint    = "0710.3820",
	archivePrefix = "arXiv",
	primaryClass  =  "hep-ph",
	doi       = "10.1016/j.cpc.2008.01.036",
	SLACcitation  = "%%CITATION = 0710.3820;%%"
}

@article{Sjostrand:2014zea,
      author         = "Sj{\"o}strand, Torbj{\"o}rn and Ask, Stefan and Christiansen,
                        Jesper R. and Corke, Richard and Desai, Nishita and Ilten,
                        Philip and Mrenna, Stephen and Prestel, Stefan and
                        Rasmussen, Christine O. and Skands, Peter Z.",
      title          = "{An introduction to PYTHIA 8.2}",
      journal        = "Comput. Phys. Commun.",
      volume         = "191",
      year           = "2015",
      pages          = "159",
      doi            = "10.1016/j.cpc.2015.01.024",
      eprint         = "1410.3012",
      archivePrefix  = "arXiv",
      primaryClass   = "hep-ph",
      reportNumber   = "LU-TP-14-36, MCNET-14-22, CERN-PH-TH-2014-190,
                        FERMILAB-PUB-14-316-CD, DESY-14-178, SLAC-PUB-16122,
                        --FERMILAB-PUB-14-316-CD",
      SLACcitation   = "%%CITATION = ARXIV:1410.3012;%%"
}

@Article{Lange:2001uf,
      author         = "Lange, D. J.",
      title          = "{The EvtGen particle decay simulation package}",
      booktitle      = "{Proceedings, 7th International Conference on B physics
                        at hadron machines (BEAUTY 2000)}",
      journal        = "Nucl. Instrum. Meth. A",
      volume         = "462",
      year           = "2001",
      pages          = "152",
      doi            = "10.1016/S0168-9002(01)00089-4",
      SLACcitation   = "%%CITATION = NUIMA,A462,152;%%"
}

@Article{Frixione:2008yi,
     author    = "Frixione, Stefano and Laenen, Eric and Motylinski, Patrick and White, Chris and Webber, Bryan R.",
     title     = "{Single-top hadroproduction in association with a \(W\) boson}",
     journal   = "JHEP",
     volume    = "07",
     year      = "2008",
     pages     = "029",
     eprint    = "0805.3067",
     archivePrefix = "arXiv",
     primaryClass  =  "hep-ph",
     doi       = "10.1088/1126-6708/2008/07/029"
}

@Article{Alwall:2014hca,
      author         = "Alwall, J. and Frederix, R. and Frixione, S. and Hirschi,
                        V. and Maltoni, F. and Mattelaer, O. and Shao, H. -S. and
                        Stelzer, T. and Torrielli, P. and Zaro, M.",
      title          = "{The automated computation of tree-level and
                        next-to-leading order differential cross sections, and
                        their matching to parton shower simulations}",
      journal        = "JHEP",
      volume         = "07",
      year           = "2014",
      pages          = "079",
      doi            = "10.1007/JHEP07(2014)079",
      eprint         = "1405.0301",
      archivePrefix  = "arXiv",
      primaryClass   = "hep-ph",
      reportNumber   = "CERN-PH-TH-2014-064, CP3-14-18, LPN14-066, MCNET-14-09,
                        ZU-TH-14-14",
      SLACcitation   = "%%CITATION = ARXIV:1405.0301;%%"
}

@Article{Bahr:2008pv,
      author         = "B{\"a}hr, M. and others",
      title          = "{Herwig++ physics and manual}",
      journal        = "Eur. Phys. J. C",
      volume         = "58",
      year           = "2008",
      pages          = "639",
      doi            = "10.1140/epjc/s10052-008-0798-9",
      eprint         = "0803.0883",
      archivePrefix  = "arXiv",
      primaryClass   = "hep-ph",
      reportNumber   = "CERN-PH-TH-2008-038, CAVENDISH-HEP-08-03, KA-TP-05-2008,
                        DCPT-08-22, IPPP-08-11, CP3-08-05",
      SLACcitation   = "%%CITATION = ARXIV:0803.0883;%%"
}

@Article{Bellm:2015jjp,
      author         = "Bellm, Johannes and others",
      title          = "{Herwig 7.0/Herwig++ 3.0 release note}",
      journal        = "Eur. Phys. J. C",
      volume         = "76",
      year           = "2016",
      number         = "4",
      pages          = "196",
      doi            = "10.1140/epjc/s10052-016-4018-8",
      eprint         = "1512.01178",
      archivePrefix  = "arXiv",
      primaryClass   = "hep-ph",
      reportNumber   = "CERN-PH-TH-2015-289, MAN-HEP-2015-15, IFJPAN-IV-2015-13,
                        HERWIG-2015-01, KA-TP-18-2015, DCPT-15-142, MCNET-15-28,
                        IPPP-15-71, --HERWIG-2015-01",
      SLACcitation   = "%%CITATION = ARXIV:1512.01178;%%"
}

@Article{Nason:2004rx,
      author         = "Nason, Paolo",
      title          = "{A new method for combining NLO QCD with shower Monte Carlo algorithms}",
      journal        = "JHEP",
      volume         = "11",
      pages          = "040",
      doi            = "10.1088/1126-6708/2004/11/040",
      year           = "2004",
      eprint         = "hep-ph/0409146",
      archivePrefix  = "arXiv",
}

@Article{Frixione:2007vw,
      author         = "Frixione, Stefano and Nason, Paolo and Oleari, Carlo",
      title          = "{Matching NLO QCD computations with parton shower
                        simulations: the POWHEG method}",
      journal        = "JHEP",
      volume         = "11",
      pages          = "070",
      doi            = "10.1088/1126-6708/2007/11/070",
      year           = "2007",
      eprint         = "0709.2092",
      archivePrefix  = "arXiv",
      primaryClass   = "hep-ph",
}

@Article{Alioli:2010xd,
      author         = "Alioli, Simone and Nason, Paolo and Oleari, Carlo and Re,
                        Emanuele",
      title          = "{A general framework for implementing NLO calculations in
                        shower Monte Carlo programs: the POWHEG BOX}",
      journal        = "JHEP",
      volume         = "06",
      pages          = "043",
      doi            = "10.1007/JHEP06(2010)043",
      year           = "2010",
      eprint         = "1002.2581",
      archivePrefix  = "arXiv",
      primaryClass   = "hep-ph",
}

@article{Frixione:2007nw,
      author         = "Frixione, Stefano and Ridolfi, Giovanni and Nason, Paolo",
      title          = "{A positive-weight next-to-leading-order Monte Carlo for
                        heavy flavour hadroproduction}",
      journal        = "JHEP",
      volume         = "09",
      pages          = "126",
      doi            = "10.1088/1126-6708/2007/09/126",
      year           = "2007",
      eprint         = "0707.3088",
      archivePrefix  = "arXiv",
      primaryClass   = "hep-ph",
      SLACcitation   = "%%CITATION = ARXIV:0707.3088;%%",
}

@article{deFlorian:2016spz,
      author         = "de Florian, D. and others",
      title          = "{Handbook of LHC Higgs Cross Sections: 4. Deciphering the
                        Nature of the Higgs Sector}",
      collaboration  = "LHC Higgs Cross Section Working Group",
      doi            = "10.23731/CYRM-2017-002",
      year           = "2016",
      eprint         = "1610.07922",
      archivePrefix  = "arXiv",
      primaryClass   = "hep-ph",
      reportNumber   = "FERMILAB-FN-1025-T, CERN-2017-002-M",
      SLACcitation   = "%%CITATION = ARXIV:1610.07922;%%"
}

@Article{Gleisberg:2008ta,
      author         = "Gleisberg, T. and H{\"o}che, S. and Krauss, F. and Sch{\"o}nherr, M. and Schumann, S. and Siegert, F. and Winter J.",
      title          = "{Event generation with SHERPA 1.1}",
      journal        = "JHEP",
      volume         = "02",
      pages          = "007",
      doi            = "10.1088/1126-6708/2009/02/007",
      year           = "2009",
      eprint         = "0811.4622",
      archivePrefix  = "arXiv",
      primaryClass   = "hep-ph",
}

@Article{Hoeche:2009rj,
      author         = "H{\"o}che, Stefan and Krauss, Frank and Schumann, Steffen and Siegert, Frank",
      title          = "{QCD matrix elements and truncated showers}",
      journal        = "JHEP",
      volume         = "05",
      pages          = "053",
      doi            = "10.1088/1126-6708/2009/05/053",
      year           = "2009",
      eprint         = "0903.1219",
      archivePrefix  = "arXiv",
      primaryClass   = "hep-ph",
}

@Article{Gleisberg:2008fv,
      author         = "Gleisberg, Tanju and H{\"o}che, Stefan",
      title          = "{Comix, a new matrix element generator}",
      journal        = "JHEP",
      volume         = "12",
      pages          = "039",
      doi            = "10.1088/1126-6708/2008/12/039",
      year           = "2008",
      eprint         = "0808.3674",
      archivePrefix  = "arXiv",
      primaryClass   = "hep-ph",
}

@Article{Schumann:2007mg,
      author         = "Schumann, Steffen and Krauss, Frank",
      title          = "{A parton shower algorithm based on Catani--Seymour dipole factorisation}",
      journal        = "JHEP",
      volume         = "03",
      pages          = "038",
      doi            = "10.1088/1126-6708/2008/03/038",
      year           = "2008",
      eprint         = "0709.1027",
      archivePrefix  = "arXiv",
      primaryClass   = "hep-ph",
}

@article{Hoeche:2012yf,
      author         = "H{\"o}che, Stefan and Krauss, Frank and Sch{\"o}nherr, Marek and
                        Siegert, Frank",
      title          = "{QCD matrix elements + parton showers. The NLO case}",
      journal        = "JHEP",
      volume         = "04",
      year           = "2013",
      pages          = "027",
      doi            = "10.1007/JHEP04(2013)027",
      eprint         = "1207.5030",
      archivePrefix  = "arXiv",
      primaryClass   = "hep-ph",
      reportNumber   = "SLAC-PUB-15191, IPPP-12-52, DCPT-12-104, LPN12-081,
                        FR-PHENO-2012-017, MCNET-12-09, --FR-PHENO-2012-017",
      SLACcitation   = "%%CITATION = ARXIV:1207.5030;%%"
}

@article{Hoeche:2011fd,
      author         = "H{\"o}che, Stefan and Krauss, Frank and Sch{\"o}nherr, Marek and
                        Siegert, Frank",
      title          = "{A critical appraisal of NLO+PS matching methods}",
      journal        = "JHEP",
      volume         = "09",
      year           = "2012",
      pages          = "049",
      doi            = "10.1007/JHEP09(2012)049",
      eprint         = "1111.1220",
      archivePrefix  = "arXiv",
      primaryClass   = "hep-ph",
      reportNumber   = "SLAC-PUB-14661, IPPP-11-67, DCPT-11-134, LPN11-58,
                        FR-PHENO-2011-019, MCNET-11-24",
      SLACcitation   = "%%CITATION = ARXIV:1111.1220;%%"
}

@article{Catani:2001cc,
      author         = "Catani, S. and Krauss, F. and Webber, B. R. and Kuhn, R.",
      title          = "{QCD Matrix Elements + Parton Showers}",
      journal        = "JHEP",
      volume         = "11",
      year           = "2001",
      pages          = "063",
      doi            = "10.1088/1126-6708/2001/11/063",
      eprint         = "hep-ph/0109231",
      archivePrefix  = "arXiv",
      reportNumber   = "CERN-TH-2000-367, CAVENDISH-HEP-00-03",
      SLACcitation   = "%%CITATION = HEP-PH/0109231;%%"
}

@article{Agostinelli:2002hh,
      author         = "{GEANT4 Collaboration} and Agostinelli, S. and others",
      title          = "{\textsc{Geant4} -- a simulation toolkit}",
      journal        = "Nucl. Instrum. Meth. A",
      volume         = "506",
      year           = "2003",
      pages          = "250",
      doi            = "10.1016/S0168-9002(03)01368-8",
      reportNumber   = "SLAC-PUB-9350, FERMILAB-PUB-03-339",
      SLACcitation   = "%%CITATION = NUIMA,A506,250;%%"
}

@Article{Read:2002hq,
    Author = {Read, Alexander L.},
    Journal = {J. Phys. G},
    Pages = {2693},
    doi = "10.1088/0954-3899/28/10/313",
    Title = "{Presentation of search results: the \(CL_S\) technique}",
    Volume = {28},
    Year = {2002}
}

@Article{Cowan:2010js,
   Author = {Cowan, Glen and Cranmer, Kyle and Gross, Eilam and Vitells, Ofer},
   Title = {{Asymptotic formulae for likelihood-based tests of new physics}},
   Journal = {Eur. Phys. J. C},
   Volume = {71},
   Year = {2011},
   Pages = {1554},
   doi = "10.1140/epjc/s10052-011-1554-0",
   Eprint = "1007.1727",
   Archiveprefix = {arXiv},
   Primaryclass = {physics.data-an},
   related = "Cowan:2010js-err",
   relatedstring  = "Erratum:",
}

@Booklet{RooFit,
      author         = "Verkerke, Wouter and Kirkby, David",
      title          = "{The RooFit toolkit for data modeling}",
      year           = "2003",
      eprint         = "physics/0306116",
      archivePrefix  = "arXiv",
      primaryClass   = "physics.data-an",
}

@Article{PERF-2007-01,
    author         = "{ATLAS Collaboration}",
    title          = "{The ATLAS Experiment at the CERN Large Hadron Collider}",
    journal        = "JINST",
    volume         = "3",
    year           = "2008",
    pages          = "S08003",
    doi            = "10.1088/1748-0221/3/08/S08003",
    primaryClass   = "hep-ex",
}

@Article{SOFT-2010-01,
    author         = "{ATLAS Collaboration}",
    title          = "{The ATLAS Simulation Infrastructure}",
    journal        = "Eur. Phys. J. C",
    volume         = "70",
    year           = "2010",
    pages          = "823",
    doi            = "10.1140/epjc/s10052-010-1429-9",
    eprint         = "1005.4568",
    archivePrefix  = "arXiv",
    primaryClass   = "physics.ins-det",
}

@Article{PERF-2011-03,
    author         = "{ATLAS Collaboration}",
    title          = "{Jet energy measurement with the ATLAS detector in proton--proton collisions at \(\sqrt{s} = 7\,\text{TeV}\)}",
    journal        = "Eur. Phys. J. C",
    volume         = "73",
    year           = "2013",
    pages          = "2304",
    doi            = "10.1140/epjc/s10052-013-2304-2",
    reportNumber   = "CERN-PH-EP-2011-191",
    eprint         = "1112.6426",
    archivePrefix  = "arXiv",
    primaryClass   = "hep-ex",
}

@Article{HIGG-2012-27,
    author         = "{ATLAS Collaboration}",
    title          = "{Observation of a new particle in the search for the Standard Model Higgs boson with the ATLAS detector at the LHC}",
    journal        = "Phys. Lett. B",
    volume         = "716",
    year           = "2012",
    pages          = "1",
    doi            = "10.1016/j.physletb.2012.08.020",
    reportNumber   = "CERN-PH-EP-2012-218",
    eprint         = "1207.7214",
    archivePrefix  = "arXiv",
    primaryClass   = "hep-ex",
}

@Article{PERF-2014-03,
    author         = "{ATLAS Collaboration}",
    title          = "{Performance of pile-up mitigation techniques for jets in \(pp\) collisions at \(\sqrt{s} = 8\,\text{TeV}\) using the ATLAS detector}",
    journal        = "Eur. Phys. J. C",
    volume         = "76",
    year           = "2016",
    pages          = "581",
    doi            = "10.1140/epjc/s10052-016-4395-z",
    reportNumber   = "CERN-PH-EP-2015-206",
    eprint         = "1510.03823",
    archivePrefix  = "arXiv",
    primaryClass   = "hep-ex",
}

@Article{EXOT-2015-04,
    author         = "{ATLAS Collaboration}",
    title          = "{Search for heavy particles decaying into top-quark pairs using lepton-plus-jets events in proton--proton collisions at \(\sqrt{s} = 13\,\text{TeV}\) with the ATLAS detector}",
    journal        = "Eur. Phys. J. C",
    volume         = "78",
    year           = "2018",
    pages          = "565",
    doi            = "10.1140/epjc/s10052-018-5995-6",
    reportNumber   = "CERN-EP-2018-48",
    eprint         = "1804.10823",
    archivePrefix  = "arXiv",
    primaryClass   = "hep-ex",
}

@Article{PERF-2015-09,
    author         = "{ATLAS Collaboration}",
    title          = "{Jet reconstruction and performance using particle flow with the ATLAS Detector}",
    journal        = "Eur. Phys. J. C",
    volume         = "77",
    year           = "2017",
    pages          = "466",
    doi            = "10.1140/epjc/s10052-017-5031-2",
    reportNumber   = "CERN-EP-2017-024",
    eprint         = "1703.10485",
    archivePrefix  = "arXiv",
    primaryClass   = "hep-ex",
}

@Article{PERF-2015-10,
    author         = "{ATLAS Collaboration}",
    title          = "{Muon reconstruction performance of the ATLAS detector in proton--proton collision data at \(\sqrt{s} = 13\,\text{TeV}\)}",
    journal        = "Eur. Phys. J. C",
    volume         = "76",
    year           = "2016",
    pages          = "292",
    doi            = "10.1140/epjc/s10052-016-4120-y",
    reportNumber   = "CERN-EP-2016-033",
    eprint         = "1603.05598",
    archivePrefix  = "arXiv",
    primaryClass   = "hep-ex",
}

@Article{EXOT-2016-13,
    author         = "{ATLAS Collaboration}",
    title          = "{Search for pair production of up-type vector-like quarks and for four-top-quark events in final states with multiple \(b\)-jets with the ATLAS detector}",
    journal        = "JHEP",
    volume         = "07",
    year           = "2018",
    pages          = "089",
    doi            = "10.1007/JHEP07(2018)089",
    reportNumber   = "CERN-EP-2018-031",
    eprint         = "1803.09678",
    archivePrefix  = "arXiv",
    primaryClass   = "hep-ex",
}

@Article{EXOT-2016-16,
    author         = "{ATLAS Collaboration}",
    title          = "{Search for new phenomena in events with same-charge leptons and \(b\)-jets in \(pp\) collisions at \(\sqrt{s} = 13\,\text{TeV}\) with the ATLAS detector}",
    journal        = "JHEP",
    volume         = "12",
    year           = "2018",
    pages          = "039",
    doi            = "10.1007/JHEP12(2018)039",
    reportNumber   = "CERN-EP-2018-171",
    eprint         = "1807.11883",
    archivePrefix  = "arXiv",
    primaryClass   = "hep-ex",
}

@Article{EXOT-2016-24,
    author         = "{ATLAS Collaboration}",
    title          = "{Search for heavy particles decaying into a top-quark pair in the fully hadronic final state in \(pp\) collisions at \(\sqrt{s} = 13\,\text{TeV}\) with the ATLAS detector}",
    journal        = "Phys. Rev. D",
    volume         = "99",
    year           = "2019",
    pages          = "092004",
    doi            = "10.1103/PhysRevD.99.092004",
    reportNumber   = "CERN-EP-2018-350",
    eprint         = "1902.10077",
    archivePrefix  = "arXiv",
    primaryClass   = "hep-ex",
}

@Article{TRIG-2016-01,
    author         = "{ATLAS Collaboration}",
    title          = "{Performance of the ATLAS trigger system in 2015}",
    journal        = "Eur. Phys. J. C",
    volume         = "77",
    year           = "2017",
    pages          = "317",
    doi            = "10.1140/epjc/s10052-017-4852-3",
    reportNumber   = "CERN-EP-2016-241",
    eprint         = "1611.09661",
    archivePrefix  = "arXiv",
    primaryClass   = "hep-ex",
}

@Article{TOPQ-2017-12,
    author         = "{ATLAS Collaboration}",
    title          = "{Measurements of inclusive and differential fiducial cross-sections of \(t\bar{t}\) production with additional heavy-flavour jets in proton--proton collisions at \(\sqrt{s} = 13\,\text{TeV}\) with the ATLAS detector}",
    journal        = "JHEP",
    volume         = "04",
    year           = "2019",
    pages          = "046",
    doi            = "10.1007/JHEP04(2019)046",
    reportNumber   = "CERN-EP-2018-276",
    eprint         = "1811.12113",
    archivePrefix  = "arXiv",
    primaryClass   = "hep-ex",
}

@Article{EGAM-2018-01,
    author         = "{ATLAS Collaboration}",
    title          = "{Electron and photon performance measurements with the ATLAS detector using the 2015--2017 LHC proton--proton collision data}",
    journal        = "JINST",
    volume         = "14",
    year           = "2019",
    pages          = "P12006",
    doi            = "10.1088/1748-0221/14/12/P12006",
    reportNumber   = "CERN-EP-2019-145",
    eprint         = "1908.00005",
    archivePrefix  = "arXiv",
    primaryClass   = "hep-ex",
}

@Article{EXOT-2018-48,
    author         = "{ATLAS Collaboration}",
    title          = "{Search for  \(t\bar{t}\) resonances in fully hadronic final states in \(pp\) collisions at \(\sqrt{s} = 13\,\text{TeV}\) with the ATLAS detector}",
    journal        = "JHEP",
    volume         = "10",
    year           = "2020",
    pages          = "061",
    doi            = "10.1007/JHEP10(2020)061",
    reportNumber   = "CERN-EP-2020-055",
    eprint         = "2005.05138",
    archivePrefix  = "arXiv",
    primaryClass   = "hep-ex",
}

@Article{FTAG-2018-01,
    author         = "{ATLAS Collaboration}",
    title          = "{ATLAS \(b\)-jet identification performance and efficiency measurement with \(t\bar{t}\) events in \(pp\) collisions at \(\sqrt{s} = 13\,\text{TeV}\)}",
    journal        = "Eur. Phys. J. C",
    volume         = "79",
    year           = "2019",
    pages          = "970",
    doi            = "10.1140/epjc/s10052-019-7450-8",
    reportNumber   = "CERN-EP-2019-132",
    eprint         = "1907.05120",
    archivePrefix  = "arXiv",
    primaryClass   = "hep-ex",
}

@Article{HIGG-2018-57,
    author         = "{ATLAS Collaboration}",
    title          = "{Combined measurements of Higgs boson production and decay using up to \(80\,\text{fb}^{-1}\) of proton--proton collision data at \(\sqrt{s} = 13\,\text{TeV}\) collected with the ATLAS experiment}",
    journal        = "Phys. Rev. D",
    volume         = "101",
    year           = "2020",
    pages          = "012002",
    doi            = "10.1103/PhysRevD.101.012002",
    reportNumber   = "CERN-EP-2019-097",
    eprint         = "1909.02845",
    archivePrefix  = "arXiv",
    primaryClass   = "hep-ex",
}

@Article{TOPQ-2018-05,
    author         = "{ATLAS Collaboration}",
    title          = "{Evidence for \(t\bar{t}t\bar{t}\) production in the multilepton final state in proton--proton collisions at \(\sqrt{s} = 13\text {TeV}\) with the ATLAS detector}",
    journal        = "Eur. Phys. J. C",
    volume         = "80",
    year           = "2020",
    pages          = "1085",
    doi            = "10.1140/epjc/s10052-020-08509-3",
    reportNumber   = "CERN-EP-2020-111",
    eprint         = "2007.14858",
    archivePrefix  = "arXiv",
    primaryClass   = "hep-ex",
}

@Article{TOPQ-2018-15,
    author         = "{ATLAS Collaboration}",
    title          = "{Measurements of top-quark pair differential and double-differential cross-sections in the \(\ell\)+jets channel with \(pp\) collisions at \(\sqrt{s} = 13\,\text{TeV}\) using the ATLAS detector}",
    journal        = "Eur. Phys. J. C",
    volume         = "79",
    year           = "2019",
    pages          = "1028",
    doi            = "10.1140/epjc/s10052-019-7525-6",
    reportNumber   = "CERN-EP-2019-149",
    eprint         = "1908.07305",
    archivePrefix  = "arXiv",
    primaryClass   = "hep-ex",
    related        = "TOPQ-2018-15-err",
    relatedstring  = "Erratum:",
}

@Article{TRIG-2018-01,
    author         = "{ATLAS Collaboration}",
    title          = "{Performance of the ATLAS muon triggers in Run~2}",
    journal        = "JINST",
    volume         = "15",
    year           = "2020",
    pages          = "P09015",
    doi            = "10.1088/1748-0221/15/09/p09015",
    reportNumber   = "CERN-EP-2020-031",
    eprint         = "2004.13447",
    archivePrefix  = "arXiv",
    primaryClass   = "hep-ex",
}

@Article{TRIG-2018-05,
    author         = "{ATLAS Collaboration}",
    title          = "{Performance of electron and photon triggers in ATLAS during LHC Run~2}",
    journal        = "Eur. Phys. J. C",
    volume         = "80",
    year           = "2020",
    pages          = "47",
    doi            = "10.1140/epjc/s10052-019-7500-2",
    reportNumber   = "CERN-EP-2019-169",
    eprint         = "1909.00761",
    archivePrefix  = "arXiv",
    primaryClass   = "hep-ex",
}

@Article{EXOT-2019-03,
    author         = "{ATLAS Collaboration}",
    title          = "{Search for new resonances in mass distributions of jet pairs using \(139\,\text{fb}^{-1}\) of \(pp\) collisions at \(\sqrt{s} = 13\,\text{TeV}\) with the ATLAS detector}",
    journal        = "JHEP",
    volume         = "03",
    year           = "2020",
    pages          = "145",
    doi            = "10.1007/JHEP03(2020)145",
    reportNumber   = "CERN-EP-2019-162",
    eprint         = "1910.08447",
    archivePrefix  = "arXiv",
    primaryClass   = "hep-ex",
}

@Article{FTAG-2020-08,
    author         = "{ATLAS Collaboration}",
    title          = "{Measurement of the \(c\)-jet mistagging efficiency in \(t\bar{t}\) events using \(pp\) collision data at \(\sqrt{s} = 13\,\text{TeV}\) collected with the ATLAS detector}",
    journal        = "Eur. Phys. J. C",
    volume         = "82",
    year           = "2021",
    pages          = "95",
    doi            = "10.1140/epjc/s10052-021-09843-w",
    reportNumber   = "CERN-EP-2021-166",
    eprint         = "2109.10627",
    archivePrefix  = "arXiv",
    primaryClass   = "hep-ex",
}

@Article{TOPQ-2020-10,
    author         = "{ATLAS Collaboration}",
    title          = "{Measurement of the \(t\bar{t}t\bar{t}\) production cross section in \(pp\) collisions at \(\sqrt{s} = 13\,\text{TeV}\) with the ATLAS detector}",
    journal        = "JHEP",
    volume         = "11",
    year           = "2021",
    pages          = "118",
    doi            = "10.1007/JHEP11(2021)118",
    reportNumber   = "CERN-EP-2021-075",
    eprint         = "2106.11683",
    archivePrefix  = "arXiv",
    primaryClass   = "hep-ex",
}

@Booklet{ATL-SOFT-PUB-2021-001,
    author         = "{ATLAS Collaboration}",
    title          = "{The ATLAS Collaboration Software and Firmware}",
    howpublished   = "{ATL-SOFT-PUB-2021-001}",
    url            = "https://cds.cern.ch/record/2767187",
    year           = "2021",
}

@Booklet{ATL-SOFT-PUB-2021-003,
    author         = "{ATLAS Collaboration}",
    title          = "{ATLAS Computing Acknowledgements}",
    howpublished   = "{ATL-SOFT-PUB-2021-003}",
    url            = "https://cds.cern.ch/record/2776662",
    year           = "2021",
}

@Article{CMS-HIG-12-028,
    author         = "{CMS Collaboration}",
    title          = "{Observation of a new boson at a mass of 125 GeV with the CMS experiment at the LHC}",
    journal        = "Phys. Lett. B",
    volume         = "716",
    year           = "2012",
    pages          = "30",
    doi            = "10.1016/j.physletb.2012.08.021",
    reportNumber   = "CERN-PH-EP-2012-220",
    eprint         = "1207.7235",
    archivePrefix  = "arXiv",
    primaryClass   = "hep-ex",
}

@Article{CMS-B2G-16-015,
    author         = "{CMS Collaboration}",
    title          = "{Search for \(t\bar{t}\) resonances in highly-boosted lepton+jets and fully hadronic final states in proton--proton collisions at \(\sqrt{s} = 13\,\text{TeV}\)}",
    journal        = "JHEP",
    volume         = "07",
    year           = "2017",
    pages          = "001",
    doi            = "10.1007/JHEP07(2017)001",
    reportNumber   = "CERN-EP-2017-049",
    eprint         = "1704.03366",
    archivePrefix  = "arXiv",
    primaryClass   = "hep-ex",
}

@Article{CMS-B2G-17-017,
    author         = "{CMS Collaboration}",
    title          = "{Search for resonant \(t\bar{t}\) production in proton--proton collisions at \(\sqrt{s} = 13\,\text{TeV}\)}",
    journal        = "JHEP",
    volume         = "04",
    year           = "2019",
    pages          = "031",
    doi            = "10.1007/JHEP04(2019)031",
    reportNumber   = "CERN-EP-2018-254",
    eprint         = "1810.05905",
    archivePrefix  = "arXiv",
    primaryClass   = "hep-ex",
}

@Article{CMS-TOP-18-003,
    author         = "{CMS Collaboration}",
    title          = "{Search for production of four top quarks in final states with same-sign or multiple leptons in proton--proton collisions at \(\sqrt{s} = 13\,\text{TeV}\)}",
    journal        = "Eur. Phys. J. C",
    volume         = "80",
    year           = "2020",
    pages          = "75",
    doi            = "10.1140/epjc/s10052-019-7593-7",
    reportNumber   = "CERN-EP-2019-163",
    eprint         = "1908.06463",
    archivePrefix  = "arXiv",
    primaryClass   = "hep-ex",
}

@Article{CMS-TOP-18-011,
    author         = "{CMS Collaboration}",
    title          = "{Measurement of the \(t\bar{t} b\bar{b}\) production cross section in the all-jet final state in \(pp\) collisions at \(\sqrt{s} = 13\,\text{TeV}\)}",
    journal        = "Phys. Lett. B",
    volume         = "803",
    year           = "2020",
    pages          = "135285",
    doi            = "10.1016/j.physletb.2020.135285",
    reportNumber   = "CERN-EP-2019-183",
    eprint         = "1909.05306",
    archivePrefix  = "arXiv",
    primaryClass   = "hep-ex",
}

@Booklet{ATLAS-CONF-2014-018,
    author         = "{ATLAS Collaboration}",
    title          = "{Tagging and suppression of pileup jets with the ATLAS detector}",
    howpublished   = "{ATLAS-CONF-2014-018}",
    url            = "https://cds.cern.ch/record/1700870",
    year           = "2014",
}

@Booklet{ATLAS-CONF-2017-062,
    author         = "{ATLAS Collaboration}",
    title          = "{Jet reclustering and close-by effects in ATLAS Run~2}",
    howpublished   = "{ATLAS-CONF-2017-062}",
    url            = "https://cds.cern.ch/record/2275649",
    year           = "2017",
}

@Booklet{ATLAS-CONF-2018-001,
    author         = "{ATLAS Collaboration}",
    title          = "{Measurement of \(b\)-tagging efficiency of \(c\)-jets in \(t\bar{t}\) events using a likelihood approach with the ATLAS detector}",
    howpublished   = "{ATLAS-CONF-2018-001}",
    url            = "https://cds.cern.ch/record/2306649",
    year           = "2018",
}

@Booklet{ATLAS-CONF-2019-021,
    author         = "{ATLAS Collaboration}",
    title          = "{Luminosity determination in \(pp\) collisions at \(\sqrt{s} = 13\,\text{TeV}\) using the ATLAS detector at the LHC}",
    howpublished   = "{ATLAS-CONF-2019-021}",
    url            = "https://cds.cern.ch/record/2677054",
    year           = "2019",
}

@Booklet{ATL-PHYS-PUB-2010-013,
    author         = "{ATLAS Collaboration}",
    title          = "{The simulation principle and performance of the ATLAS fast calorimeter simulation FastCaloSim}",
    howpublished   = "{ATL-PHYS-PUB-2010-013}",
    url            = "https://cds.cern.ch/record/1300517",
    year           = "2010",
}

@Booklet{ATL-PHYS-PUB-2012-003,
    author         = "{ATLAS Collaboration}",
    title          = "{Summary of ATLAS Pythia 8 tunes}",
    howpublished   = "{ATL-PHYS-PUB-2012-003}",
    url            = "https://cds.cern.ch/record/1474107",
    year           = "2012",
}

@Booklet{ATL-PHYS-PUB-2014-021,
    author         = "{ATLAS Collaboration}",
    title          = "{ATLAS Pythia~8 tunes to \(7~\text{TeV}\) data}",
    howpublished   = "{ATL-PHYS-PUB-2014-021}",
    url            = "https://cds.cern.ch/record/1966419",
    year           = "2014",
}

@Booklet{ATL-PHYS-PUB-2015-015,
    author         = "{ATLAS Collaboration}",
    title          = "{Jet Calibration and Systematic Uncertainties for Jets Reconstructed in the ATLAS Detector at \(\sqrt{s} = 13~\text{TeV}\)}",
    howpublished   = "{ATL-PHYS-PUB-2015-015}",
    url            = "https://cds.cern.ch/record/2037613",
    year           = "2015",
}

@Booklet{ATL-PHYS-PUB-2015-026,
    author         = "{ATLAS Collaboration}",
    title          = "{Vertex Reconstruction Performance of the ATLAS Detector at \(\sqrt{s} = 13~\text{TeV}\)}",
    howpublished   = "{ATL-PHYS-PUB-2015-026}",
    url            = "https://cds.cern.ch/record/2037717",
    year           = "2015",
}

@Booklet{ATL-PHYS-PUB-2016-020,
    author         = "{ATLAS Collaboration}",
    title          = "{Studies on top-quark Monte Carlo modelling for Top2016}",
    howpublished   = "{ATL-PHYS-PUB-2016-020}",
    url            = "https://cds.cern.ch/record/2216168",
    year           = "2016",
}
 
\clearpage
% ATLAS Collaboration author list
% Reference date of EXOT-2022-14 is 2022-12-12
% Author list last updated on date 23-DEC-23
% Data extracted on 23-Dec-2023 for paper reference EXOT-2022-14
% at 3:45pm
 
\begin{flushleft}
\hypersetup{urlcolor=black}
{\Large The ATLAS Collaboration}

\bigskip

\AtlasOrcid[0000-0002-6665-4934]{G.~Aad}$^\textrm{\scriptsize 102}$,
\AtlasOrcid[0000-0002-5888-2734]{B.~Abbott}$^\textrm{\scriptsize 120}$,
\AtlasOrcid[0000-0002-1002-1652]{K.~Abeling}$^\textrm{\scriptsize 55}$,
\AtlasOrcid[0000-0001-5763-2760]{N.J.~Abicht}$^\textrm{\scriptsize 49}$,
\AtlasOrcid[0000-0002-8496-9294]{S.H.~Abidi}$^\textrm{\scriptsize 29}$,
\AtlasOrcid[0000-0002-9987-2292]{A.~Aboulhorma}$^\textrm{\scriptsize 35e}$,
\AtlasOrcid[0000-0001-5329-6640]{H.~Abramowicz}$^\textrm{\scriptsize 151}$,
\AtlasOrcid[0000-0002-1599-2896]{H.~Abreu}$^\textrm{\scriptsize 150}$,
\AtlasOrcid[0000-0003-0403-3697]{Y.~Abulaiti}$^\textrm{\scriptsize 117}$,
\AtlasOrcid[0000-0003-0762-7204]{A.C.~Abusleme~Hoffman}$^\textrm{\scriptsize 137a}$
% ...
\bigskip
\\

% ...
$^{29}$Physics Department, Brookhaven National Laboratory, Upton NY; United States of America.\\
% ...
$^{35}$$^{(a)}$Facult\'e des Sciences Ain Chock, R\'eseau Universitaire de Physique des Hautes Energies - Universit\'e Hassan II, Casablanca;$^{(b)}$Facult\'{e} des Sciences, Universit\'{e} Ibn-Tofail, K\'{e}nitra;$^{(c)}$Facult\'e des Sciences Semlalia, Universit\'e Cadi Ayyad, LPHEA-Marrakech;$^{(d)}$LPMR, Facult\'e des Sciences, Universit\'e Mohamed Premier, Oujda;$^{(e)}$Facult\'e des sciences, Universit\'e Mohammed V, Rabat;$^{(f)}$Institute of Applied Physics, Mohammed VI Polytechnic University, Ben Guerir; Morocco.\\
% ...
$^{49}$Fakult\"{a}t Physik , Technische Universit{\"a}t Dortmund, Dortmund; Germany.\\
% ...
$^{55}$II. Physikalisches Institut, Georg-August-Universit\"{a}t G\"ottingen, G\"ottingen; Germany.\\
% ...
$^{102}$CPPM, Aix-Marseille Universit\'e, CNRS/IN2P3, Marseille; France.\\
% ...
$^{117}$Department of Physics, New York University, New York NY; United States of America.\\
% ...
$^{120}$Homer L. Dodge Department of Physics and Astronomy, University of Oklahoma, Norman OK; United States of America.\\
% ...
$^{137}$$^{(a)}$Departamento de F\'isica, Pontificia Universidad Cat\'olica de Chile, Santiago;$^{(b)}$Millennium Institute for Subatomic physics at high energy frontier (SAPHIR), Santiago;$^{(c)}$Instituto de Investigaci\'on Multidisciplinario en Ciencia y Tecnolog\'ia, y Departamento de F\'isica, Universidad de La Serena;$^{(d)}$Universidad Andres Bello, Department of Physics, Santiago;$^{(e)}$Instituto de Alta Investigaci\'on, Universidad de Tarapac\'a, Arica;$^{(f)}$Departamento de F\'isica, Universidad T\'ecnica Federico Santa Mar\'ia, Valpara\'iso; Chile.\\
% ...
$^{150}$Department of Physics, Technion, Israel Institute of Technology, Haifa; Israel.\\
$^{151}$Raymond and Beverly Sackler School of Physics and Astronomy, Tel Aviv University, Tel Aviv; Israel.\\
% ...
$^{a}$ Also Affiliated with an institute covered by a cooperation agreement with CERN.\\
$^{b}$ Also at An-Najah National University, Nablus; Palestine.\\
$^{c}$ Also at Borough of Manhattan Community College, City University of New York, New York NY; United States of America.\\
$^{d}$ Also at Bruno Kessler Foundation, Trento; Italy.\\
$^{e}$ Also at Center for High Energy Physics, Peking University; China.\\
$^{f}$ Also at Center for Interdisciplinary Research and Innovation (CIRI-AUTH), Thessaloniki; Greece.\\
% ...

\end{flushleft}

% Created with Glance <Atlas.Glance@cern.ch>

\end{document}